# Towards European Standards for Quantum Technologies

O. van Deventer[1,§], N. Spethmann[2], M. Loeffler[3], M. Amoretti[4,24], R. van den Brink[5], N. Bruno[6,7], P. Comi[8], N. Farrugia[9], M. Gramegna[10,*], B. Kassenberg[11], W. Kozlowski[12,23], T. Länger[13], T. Lindstrom[14], V. Martin[15], N. Neumann[1], H. Papadopoulos[16], S. Pascazio[17,18], M. Peev[19], R. Pitwon[20], M.A. Rol[21], P. Traina[10], P. Venderbosch[11], F. K. Wilhelm-Mauch[22], A. Jenet[25]

[1] TNO, Nederlandse Organisatie voor Toegepast Natuurwetenschappelijk Onderzoek, Netherlands
[2] PTB, Physikalisch-Technische Bundesanstalt, Germany
[3] DIN, Deutsches Institut für Normung, Germany
[4] CINI, Consorzio Interuniversitario Nazionale per l'Informatica, Italy
[5] Delft Circuits, Netherlands
[6] CNR-INO, Consiglio Nazionale delle Ricerche - Istituto Nazionale di Ottica, Italy
[7] LENS, European Laboratory for Non-Linear Spectroscopy, Italy
[8] Italtel, Italy
[9] University of Malta, Malta
[10] INRIM, Istituto Nazionale di Ricerca Metrologica, Italy
[11] QuiX Quantum, Netherlands
[12] QuTech, Advanced research center for quantum computing & quantum internet, Netherlands
[13] StandICT.eu, Supporting European Experts Presence in International Standardisation Activities in ICT
[14] NPL, National Physical Laboratory, UK
[15] UPM, Universidad Politecnica de Madrid, Spain
[16] NCSR Demokritos, Greece
[17] Università di Bari, Italy
[18] INFN, Istituto Nazionale di fisica Nucleare, Italy
[19] Huawei Technologies Duesseldorf GmbH, Germany
[20] Resolute Photonics, UK
[21] Orange Quantum Systems, Netherlands
[22] Forschungszentrum Jülich GmbH, Germany
[23] TUDelft, Technische Universiteit Delft, Netherlands
[24] Università di Parma, Italy
[25] European Commission, Joint Research Centre (JRC), Belgium

[§] Chair of CEN CENELEC FGQT: oskar.vandeventer@tno.nl

# Abstract

The Second Quantum Revolution facilitates the engineering of new classes of sensors, communication technologies, and computers with unprecedented capabilities. Supply chains for quantum technologies are emerging, some focussed on commercially available components for enabling technologies and/or quantum-technologies research infrastructures, others with already higher technology-readiness levels, near to the market.

In 2018, the European Commission has launched its large-scale and long-term Quantum Flagship research initiative to support and foster the creation and development of a competitive European quantum technologies industry, as well as the consolidation and expansion of leadership and excellence in European quantum technology research. One of the measures to achieve an accelerated development and uptake has been identified by the Quantum Flagship in its Strategic Research Agenda: the promotion of coordinated, dedicated standardisation and certification efforts.

Standardisation is indeed of paramount importance to facilitate the growth of new technologies, and the development of efficient and effective supply chains. The harmonisation of technologies, methodologies, and interfaces enables interoperable products, innovation, and competition, all leading to structuring and hence growth of markets. As quantum technologies are maturing, time has come to start thinking about further standardisation needs.

This article presents insights on standardisation for quantum technologies from the perspective of the CEN-CENELEC Focus Group on Quantum Technologies (FGQT), which was established in June 2020 to coordinate and support the development of standards relevant for European industry and research.

# Key words

quantum technologies, standardisation, CEN-CENELEC, roadmap, review, quantum computing, quantum communication, quantum metrology, focus group on quantum technologies, FGQT

# Contents

# 1. Introduction

## 1.1 Quantum technologies

Even though quantum physics is over a century old, the last twenty years of progress in science and technology have led to a tremendous level of control over actual quantum systems at the most elementary level. It is now possible to routinely prepare, trap, manipulate and detect single quantum particles, such as artificial and actual atoms, electrons, and photons. Together with the possibility of creating and controlling distinct quantum states, such as superposition states and entangled states, this second quantum revolution facilitates engineering of new classes of sensors, communication techniques, and computers with unprecedented capabilities.

Quantum Technologies (QT) allow engineering of novel devices and infrastructures with the promise of many new applications in a number of domains that can contribute to the solution of some of today's most pressing social and economic challenges. These technologies offer capabilities beyond any classical technique. Examples include achieving higher sensitivity, lower power consumption and automatic higher security, maintenance-free quantum-referenced operation for more reliable industrial facilities, etc. Furthermore, QT paves the way for novel methods as for instance for earth surveys in times of climate change, exploration of natural resources as well as information transmission and processing, and, specifically, with respect to the last item, novel methods for unprecedented security in communication. QT-based applications are approaching the market and will be a pivotal factor for success in a wide and diverse range of industries and businesses. These technologies are vital to European independence and safety, as the fields of information processing, storage, transmission and security at large are affected by them.

We argue in this article that standardisation and mapping standardisation opportunities at a relatively early stage of the technology value chain is beneficial. We observe among research communities sometimes a reluctance to engage in standardisation activities at an early stage of the technology readiness scale. Indeed, in bridging research and innovation with markets, standardisation plays a fundamental role in the valorisation of research results and knowledge transfer, particularly for matured research fields. However, for many scientists it remains unclear how standardisation can benefit science at an early technology readiness level. The establishment of the CEN-CENELEC Focus Group on Quantum Technologies (FGQT), which was tasked with a road mapping, took place at a comparatively early moment. Quantum technologies vary stark in regard of readiness levels among the different technologies packed under the umbrella of quantum technologies, with several at a fairly low level.

This article highlights those domains in which focus group stakeholders see that standardisation activities would be beneficial. While traditionally standards are made by industry for industry, in the here described context also science benefits and is an important beneficiary of standardisation activities. Using standardisation as a mean for valorisation and knowledge transfer, researchers are motivated to perceive their research results in a technology context in which interoperability and system integration are important factors.

## 1.2 Standardisation as a catalyst for innovation

Traditionally, standardisation has often been perceived as standing in contradiction to innovation [1]. On the contrary, standardisation is one of the most adequate and powerful tools to quickly capitalise and disseminate knowledge and have it implemented in the industry. That is to transfer research results to the market. In addition, the standardisation process, as such, is a knowledge sharing and knowledge production process because it serves as a common platform for actors with heterogeneous backgrounds, capacities and knowledge, i.e. research, industry, academia, public administration, and the wider society.

The European standards developing organisations CEN-CENELEC and ETSI define a standard as "a document, established by consensus and approved by a recognised body that provides, for common and repeated use, rules, guidelines or characteristics for activities or their results, aimed at the achievement of the optimum degree of order in a given context. Standards should be based on consolidated results of science, technology and experience, and aimed at the promotion of optimum community benefits" [2].

Standards bring along a number of benefits. They enable a reduction of costs and an improvement of efficiency, they ensure the quality, safety, and security of products and/or services, and support compliance with relevant legislation including EU regulations. Standards satisfy customer expectations and requirements, enable access to markets and to customers in other countries. Standards achieve compatibility and interoperability between products and components and increase knowledge about new technologies and innovations [3].

Generally, standards are developed by groups of experts from industry and research. However, other interested parties, such as from policy and administration, or environmental or consumer protection, also participate in the development of relevant standards. The development can take place at different levels: If the experts convene on a national level, national standards are developed by a consensus-based process in national standardisation bodies (NSBs). However, the NSBs may also delegate experts to the responsible committees at European level (CEN, CENELEC) or international level (ISO, IEC), where they develop the technical content as representatives for the respective countries in consensus with other delegated experts. This makes it possible to combine the positions of different countries in one standard, securing the best possible outcome for all involved parties. Other standards developing organisations, like ETSI and ITU-T, form groups and committees from delegates of commercial companies and research organisations.

There are different types of standards, whereby the type of standards actually being developed often depends on the technological readiness level (TRL), [4], of the innovation. For example, terminology standards and measurement standards are more likely to be developed at the beginning. As the TRL increases, performance, benchmark or interface standards are more needed. At the end of the innovation, when the product has reached a certain market maturity, interface or certification standards will be necessary. This is not a rule, but it provides a basis for considering which standards can be developed in which order. Nevertheless, standardisation is a public and voluntary process. This means that anyone who has the knowledge in a specific area can be empowered by their national standardisation body to set informed standards. The actual questions of which topic to standardise and when are decided independently by the community itself in a consensus-based process.

## 1.3 CEN-CENELEC Focus Group on Quantum Technologies

In 2018, the European Commission has launched its large-scale (up to one billion euro) and long-term (10 years) Quantum Flagship research initiative to "(...) kick-start a competitive European industry in Quantum Technologies and to make Europe a dynamic and attractive region for innovative research, business and investments in this field" [5]. Its Strategic Research Agenda stresses that to achieve the goals of the Quantum Flagship "(…) it is necessary to accelerate the development and take-up by the market, which would be further enhanced through dedicated standardisation and certification efforts" [6].

In order to coordinate and support the development of relevant QT standards, the European standards developing organisation CEN-CENELEC in June 2020 kicked off its Focus Group on Quantum Technologies (FGQT), [7]. The group is developing its FGQT Standardisation Roadmap (publication planned for early 2023) to systematically address ongoing and prospective standardisation efforts. This activity evolves in conjunction with an identification of relevant use cases, potential QT-related transactions and supply chains, and specifically includes an analysis of aspects of QTs that would benefit most from standardisation, and within which time frame. The FGQT has currently more than 100 members from industry, research, and administration, and operates on a European level. Nevertheless, it aims at interaction with other standards developing organisations and QT-alliances world-wide, including ETSI, ITU-T, ISO/IEC, IEEE, IRTF, QuIC, etc. Another objective of the FGQT is the definition of terms-of-reference that would trigger the actual standards development in technical committees. The authors of this article are all delegates and contributors to the FGQT.

## 1.4 This review article

This article is a review of insights gained by the FGQT during its work on standardisation relevant to quantum technologies.

Section 2 and Annex A review related work, including ongoing and planned standardisation and pre-standardisation activities by other standards developing organisations (SDOs) and industry forums. Section 3 provides a view on how the FGQT aims to structure the standardisation discussions. Next, it provides initial analyses in the areas of quantum communication, quantum computing & simulation, and quantum metrology, sensing & imaging. Section 4 and Annex B present views from individual FGQT delegates, explaining their rationale of why their organisations are contributing to standards work for quantum technologies. Section 5 concludes by highlighting the early-stage work-in-progress nature of standardisation for quantum technologies.

# 2. Related work

Whereas the CEN/CENELEC FGQT may likely be the first standards developing organisation (SDO) aiming at developing a standardisation roadmap for the entire spectrum of quantum technologies, it is definitely not the first organisation to address standardisation of quantum technologies. Annex A provides an overview of past and current activities in the field being undertaken by SDOs and other organisations, on a European level and beyond, as well as references to published QT standards. This section summarises this related work, see also Figure 1.

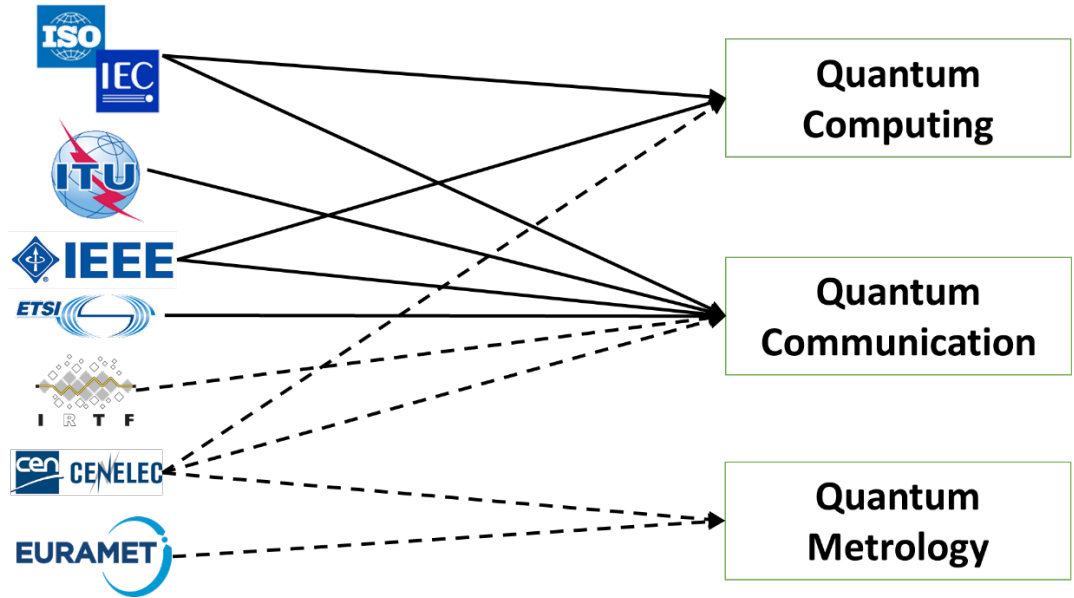

Figure 1: Standardisation activities on quantum technologies. Solid lines: standards development. Dotted lines: pre-standardisation activities.

Quantum-computing standards development has been started with ISO/IEC, focussing on quantum-computing terminology. IEEE is developing standards for quantum-computing performance metrics and technical architecture.

Quantum-communication standards are being developed by several SDOs. Both ETSI and ITU-T are telecom-industry driven SDOs. Both have a focus on quantum key distribution (QKD), which may be the first application of the quantum communication in general. Security standards for QKD are being developed by ETSI, ITU-T as well as ISO/IEC. IEEE and IRTF work on the management and control of a wider category of quantum-communication networks, including entanglement-based quantum networks and quantum internet.

Quantum-metrology standardisation is still in an emergent phase. EURAMET has recently started coordination activities for this.

Other relevant initiatives in the European context are QuIC, the European Quantum Industry Consortium, which supports European quantum-technology standardisation with industry surveys, and StandICT.eu, which provides funding to individual European ICT standardisers including some of the co-authors.

At the time of writing this article, DIN, the German National Standardisation Body (NSB) proposed to establish a European CEN-CENELEC Joint Technical Committee on Quantum Technologies, JTC22-QT, based on intermediate results from FGQT. This new technical committee will develop and coordinate European standardisation activities on quantum computing, quantum communication and quantum metrology, sensing and imaging. If the proposal is accepted, then JTC22-QT will likely be kicked-off early January 2023.

# 3. Analysing standards needs for quantum technologies

## 3.1 General

This section provides ingredients for the standardisation roadmap that FGQT is currently developing.

As mentioned in the introduction, FGQT has been analysing standards need for QT since mid 2020, and publication of a first release of its standardisation roadmap is planned early 2023. Developing a standardisation roadmap may be a somewhat chaotic process. There are many inputs to consider, including standards development elsewhere (Annex A), other relevant roadmaps like that of the European Quantum Flagship [6], and the interests of a plethora of stakeholders. All the work is carried out by volunteers in the sense that CEN-CENELEC is not paying any of the delegates for their contributions. This means the work would be seen as an investment by the organisations that provide delegates. In some cases, the own investment is augmented by European or national grants. The purpose of the roadmap is to coordinate and align interests between delegates to a point that next steps can be made, like the starting of actual standards development.

The "ingredients" to the FGQT roadmap development are of different types. Sections 3.2, 3.3, and 3.4 look at standardisation from the technological perspective for quantum computing & simulation, quantum communication, and quantum metrology, sensing & imaging, respectively. Section 3.5 looks at standardisation from the use-case perspective, discussing a selection of specific applications of quantum technologies. Section 3.6 provides an abstract market model, that aims to connect standards needs to concrete products and services, as well as to their vendors and purchasers. Section 3.7 attempts to converge the storyline of the FGQT roadmap, and to provide some structuring to the various aspects of quantum technologies.

As all of this remains work-in-progress, none of the insights and views are complete or definitive.

## 3.2 Quantum computing & simulation

Quantum Computing and Simulation as an area covers many different implementations, and several enterprises are developing solutions for a mature quantum computer. The concept of a "Modular Quantum Computer", well known from digital computing, has created a new market which has attracted many small enterprises to develop dedicated modules which are competing with more monolithic full-stack organisations. The availability of a supply chain of such modules from different vendors will enable research teams to concentrate their research on breaking new grounds, without spending much effort on duplicating known solutions. This is where standardisation can play an important role.

From a standardisation point of view, this market requires a subdivision of the field of Quantum Computing and Simulation into a variety of modules that can interwork with each other through well-defined interfaces (hardware and software), and a consensus on the functional and performance requirements of each module of interest. Instead of communicating such requirements with a single or small number of local suppliers, research teams can save effort by communicating these requirements with relevant standardisation bodies. It will increase the availability of mature hardware and software solutions in return, as well as knowledge on requirements and solutions from others.

It may be worth noting that a Quantum Simulator is a dedicated Quantum Computer, designed for solving specific problems as well as studying well defined quantum systems. They may be programmable up to a certain level. The modularity described in this paper covers both Quantum Computers and Quantum Simulators, since they use the same hardware components.

**Hardware Stack**
A first standardisation challenge that has been tackled is achieving consensus on how to subdivide the field of Quantum Computing into smaller and less-complicated chunks. A convenient way of doing that is via a stack of layers. Figure 2 below illustrates the present state of consensus within FGQT of CEN-CENELEC, covering mainly the lowest-level (hardware) layers. The layering is chosen in such a manner that their functionality can be described independently and that the interworking between different layers can be described through well-defined interfaces at their boundaries. A module from a single supplier may cover functionality within a single layer or span multiple layers. In the latter case, interfaces may be virtual (hidden internally within a module).

So far, the following low-level hardware layers have been identified:
- ***Quantum device(s)***, which may include the housing, shielding, magnets, I/O connection, etc. around one or more holders with quantum devices.
- ***Control Highway***, which may include the wiring and/or fibres, free-space optics, active devices (amplifiers), passive devices (attenuators, filters, couplers), opto-electronics (photo diodes), thermalisation means, vacuum feed-through and i/o connectors to implement a full i/o chain between quantum devices and control electronics.
- ***Control Electronics/Optics***, which may include output generators, input analysers, signal processing, i/o connection, as well as low-level firmware to guide the generation of signals and the read-out of their response.
- ***Control Software***, which may include calibration means and low-level code to translate instructions from higher software layers into commands for guiding the control electronics/optics.

The software layers above the control software layer are described further on in this section Some of the hardware layers are a mix of both hardware and software, which is illustrated via different colours in Figure 2.

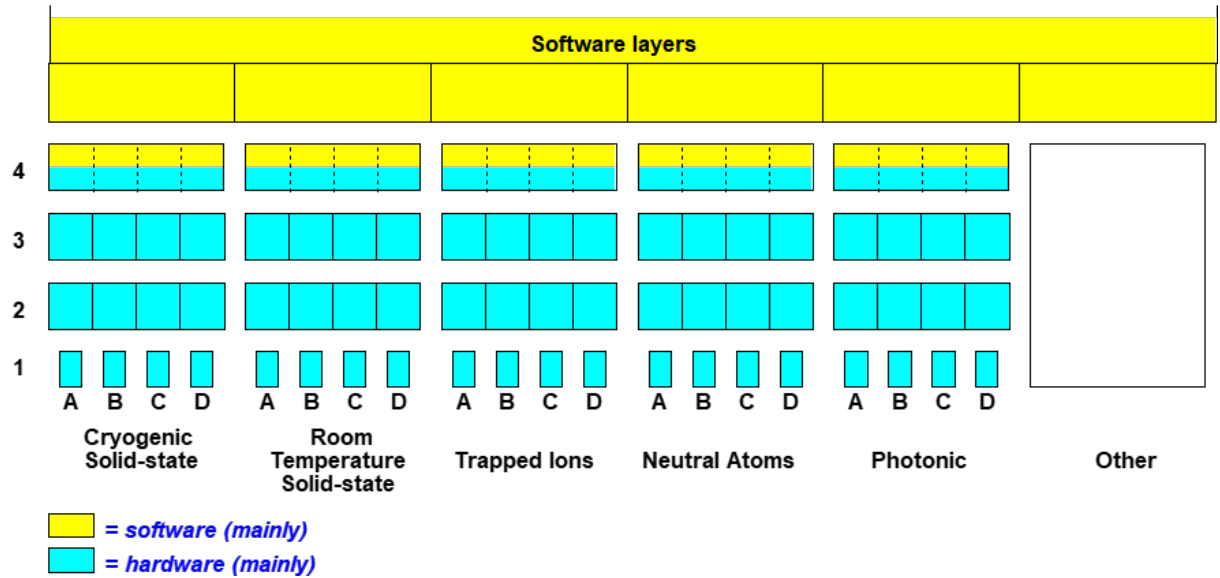

Figure 2: A possible break-down of quantum computing into layered stacks, accounting for different architectures

The layered approach allows for using different hardware stacks for specifying the requirements of dedicated architecture families. Each architecture family can have multiple members (A, B, C,...) and the description of its hardware layers (1,2,3,4) may account for differences between these members. This is illustrated symbolically via different "boxes" per layer and family, but may be merged when different members share the same functionality.
So far, the following architecture families have been identified (in arbitrary order):
*Cryogenic Solid State based*. This family covers superconducting solutions (Transmons, flux qubits), semiconductor spin qubits, topological qubits and artificial atoms in solids.
*Room Temperature Solid state based*, which cover artificial atoms in solids (such as NV Centres) and optical quantum dots.
*Trapped Ions*, covering both room-temperature and cryogenic (4K) solutions such as Optical qubits, Raman qubits and Spin (microwave) qubits.
*Neutral Atoms*, covering both collision-based and Rydberg-based solutions.
*Photonic Quantum Computing*, covering solutions based on continuous variables, cluster states / measurement based, Knill-LaFlamme-Milburn and Boson sampling.
This classification is currently work-in-progress within FGQT of CEN-CENELEC, and may be improved/revised as new insights arise.

A next step in standardisation is identifying functional requirements for modules within these layers, and there is no need to wait for the field to fully mature before starting this work. If we take Cryogenic Solid state based quantum computers as an example, one may consider the following types of requirements:
- *Quantum devices*: The modules of this architecture family are typically operating at cryogenic temperatures and may be implemented stand-alone, as chip and/or on PCB. Relevant requirements of general nature are related to shielding, materials compatibility, operating temperature, electrical and magnetic aspects, vacuum properties, and interconnectivity.
- ***Control highway***: These modules cover all infrastructure needed for routing light and microwave signals, RF and DC signals between the control electronics at room temperature and the quantum device at cryogenic temperatures. It is a mix of transmission lines, filtering, attenuation, amplification, (de)multiplexing, etc. Relevant requirements of general nature are related to low thermal conductance, good thermalisation, materials with low outgassing, vacuum feed-throughs, small footprint (for high-density i/o channels), interconnection, bandwidth, filtering properties, low noise, etc.
- ***Control electronics***: This covers all electronics for generating, receiving, and processing microwave, RF and DC signals. Some implementations make use of routing/switching and/or multiplexing of control signals at room temperatures and at cryogenic temperatures. It may also have some firmware on chip/board to route signal preparation, control, and processing/readout. Relevant requirements for consideration are dealing with signal shapes and levels, sensitivity and functionality. But also dealing with the instruction set and software interfacing with higher layers.
- ***Control software***: This covers a mix of hardware and low-level driver software for instructing the control electronics, and should provide means for calibration. Relevant requirements for consideration are dealing with the software interface to higher layers for receiving sequences of instructions about when, where and what pulses or signals are to be generated, how to process and read-out the response, and means for performing calibration.

**Software Stack**

Another standardisation challenge that has been tackled is achieving consensus on how to subdivide the software stack on top of the hardware stack. The software layers above the control software layer may include operating system, communication primitives, software drivers, hardware abstractions, assembly / register level programming, high-level programming environments and applications/services supporting use cases. The higher software layers are assumed to be more agnostic to differences in hardware.

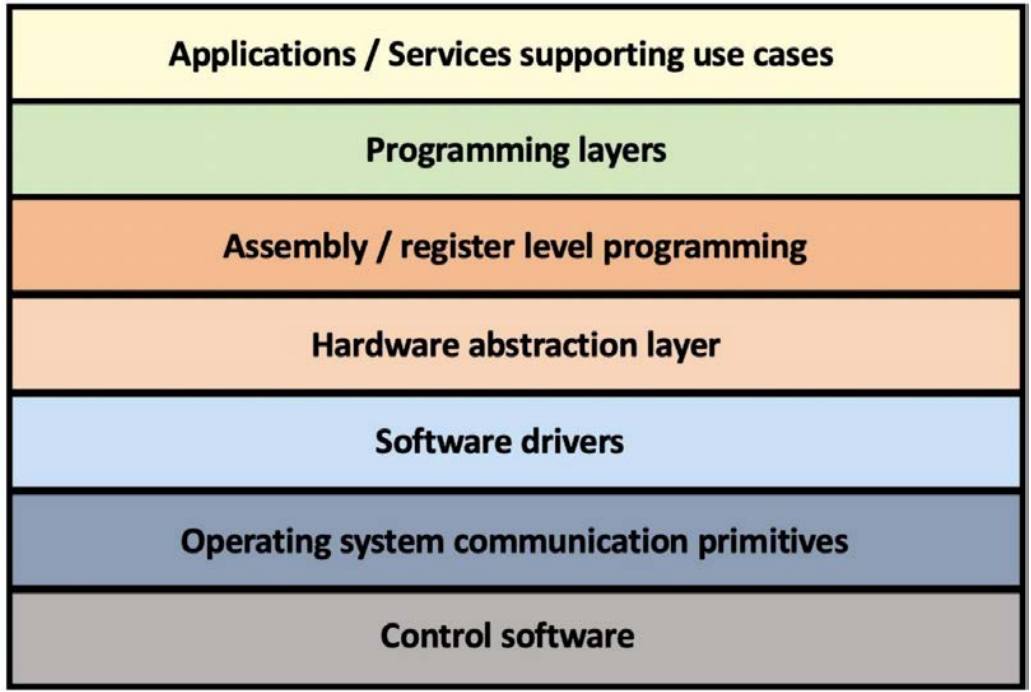

Figure 3: A possible break-down of the software stack into layers

Figure 3 illustrates the present state of consensus within FGQT of CEN-CENELEC, covering the layers in the software stack. This layering is currently work-in-progress within FGQT of CEN-CENELEC, and may also be improved/revised as new insights arise.
So far, the following software layers have been identified:

- ***Control Software***: The lowest layer in the software stack in Figure 3 (control software) is the same layer as the highest layer in the hardware stack in Figure 2. It may include calibration means, low-level code to translate instructions from higher software layers into commands for guiding the control electronics/optics, and comprises the techniques used to define error-robust physical operations and associated supporting protocols designed to tune-up and stabilize the hardware.
Control software for quantum hardware is typically stored on digital computers, i.e. there is a very strict separation between the place where the control software is stored and the quantum registers.
In the long term, control software may work in concert with Quantum Error Correction (QEC), which is supposed to lay at the assembly / register level programming layer, to provide broad coverage of various error types. More specifically, control software could improve the efficiency of QEC, i.e., reduce resource overheads required for encoding, by homogenising error rates and reducing error correlations.
- ***Operating system communication systems***: A quantum computer must be provided with an operating system (OS), which is a resource manager for the underlying quantum hardware, provided with built-in networking functions allowing multiple users and applications to use the resources as remote clients. To an application, it appears as if it has its own resources and is protected from other applications. Applications can make use of facilities only as offered by the OS. For example, the OS provides communication primitives (e.g., based on the POSIX standard for the sockets interface [8]) and only by means of these primitives should it be possible to pass messages between client applications and the quantum computer.

- *Software drivers:* These are components that are plugged into the operating system and allow hardware-abstraction programs to call the control software of the underlying quantum hardware. If the hardware changes, the software drivers must change as well.
- *Hardware Abstraction Layer (HAL):* The hardware abstraction layer (HAL) should allow quantum computer users, such as application developers, platform and system software engineers and cross-platform software architects, to abstract away the quantum computer implementation details while keeping the performance. The hardware may change, but the QASM-like programs (belonging to the upper assembly / register level programming layer) should still be able to work.
  
  Among all software layers for quantum computing, this is the one that requires the most urgent standardisation effort. The hardware abstraction layer should provide Application Programming Interfaces (APIs) to the upper layer, decoupling from the different types of quantum hardware technologies.
- *Assembly / register level programming:* This layer concerns QASM (i.e., quantum assembly) languages that describe quantum computations according to one specific model (e.g., circuit model, measurement-base model, quantum annealing model), with a per-architecture instruction set.
  
  An example is OpenQASM [9], which targets IBM Q devices and enables experiments with small depth quantum circuits. OpenQASM represents universal circuits over the CNOT plus SU(2) basis with straight-line code that includes measurement, reset, fast feedback, and gate subroutines. OpenQASM possesses a dual nature as an assembly language and as a hardware description language.
  
  A different example is NetQASM [10], which is a platform-independent and extendable universal instruction set with support for local quantum gates, digital logic, and quantum networking operations for remote entanglement generation. NetQASM consists of a specification of a low-level assembly-like language to express the quantum parts of quantum network program code.
  
  Due to the huge diversity of quantum computing architectures, it is not likely that a unique, widely accepted QASM would emerge and later become a standard.
- *Programming layers:* The specification of quantum algorithms using QASM languages is not easy for programmers. Indeed, QASM programs are usually generated by a software library, from a piece of code written in a common programming language, such as Python. In general, the programming layers include all the languages, libraries, and software development facilities (e.g., software development kits, debugging tools, quantum compilers) used by developers for coding quantum algorithms or high-level applications that use predefined quantum algorithms as subroutines.
  
  Quantum compilation is the problem of translating an input quantum circuit into the most efficient equivalent of itself, considering the characteristics of the device that will execute the computation and minimizing the number of required two-qubit gates. The most advanced quantum compilers are noise-adaptive, i.e., they take the noise statistics of the device into account.

- **Applications / Services supporting use cases**: To effectively support industrial and research use cases, quantum applications must be executed in suitable environments. Currently, some vendors provide access to quantum devices via user-friendly cloud platforms. The quantum programs must be locally compiled for a specific device and submitted for batch processing to the remote platform. However, other paradigms are emerging. For example, the Quantum Internet will enable networked quantum applications, whose execution will involve multiple quantum nodes and will be characterised by interleaved digital and quantum message passing.

## 3.3 Quantum metrology, sensing & imaging

Quantum Metrology & Sensing and quantum enhanced Imaging (QMSI) exploit the properties of quantum states and peculiar phenomena, such as entanglement and non-classical correlations, to significantly improve the accuracy and precision with which parameters of a wide range of systems can be estimated and to step over limitations related to conventional classical measurement strategies, as the environment-induced noise from vacuum fluctuations (the so-called shot noise), or the dynamically induced noise in the position measurement (the standard quantum limit), or the diffraction limit [11], [12], [13], [14].

Every aspects of the physical reality, including the measuring devices used to extract information about our real world are governed by quantum mechanics, and although this imposes unavoidable fundamental limits to each measurement process - in the specific defined by the Heisenberg uncertainty principle together with other quantum constraints on the speed of evolution - the aforementioned conventional semi-classical bounds to measurement precision are not the ultimate limits and can be beaten using suitable quantum strategies. To this purpose, these QMSI new paradigms require development of techniques robust to noise and imperfections i.e., fit to real-world scenarios, spanning from very fundamental to very practical applications, and ranging from the nanoscale, by means of localised spins to the planetary scale, based on photons.

QMSI systems and devices are very promising, and in recent years an exponential growth in the interest toward applications of these technologies was observed, and now encompasses a broad range of topics [3], [15], [16], [17], [18]. However, the diffusion in the industrial or commercial world of QMSI devices is still extremely limited. Prototypes already exist, but a mass market and standardisation of these prototypes do not. This is mainly due to the lack of reliable tools and structured facilities to characterise, test and validate the current and future prototype devices. Moreover, there is a high variety of readiness level among the different technologies in this field and the maturity level is technology dependant.

From the standardisation perspective, as stated in [3], standards in the field of QMSI are understood in different ways. Quantum Technology has enabled the discovery of new magnitudes of fundamental measurements and a redefinition of primary reference standards of the International System of Units (SI) [19], [20]. With the purpose to create a common ground of understanding, a first important aspect to be clarified is that on one side, the metrology vocabulary defines the fundamental reference standards of weight and measurements (SI), while the standardisation vocabulary generally leads to documentary standards.

In general, documentary standards represent a key step for fostering QMSI technology spreading, both for stand-alone sensors and for those to be integrated in more complex systems. The standardisation process that has just began aims to provide the way for a possible construction of a general hardware platform at a reasonable cost, a step that would represent a key enabler for these QTs, in particular in fields in which commercial product are emerging, e.g. quantum sensors based on NV-centers in diamond, atomic-based sensors, etc.

So far, the two major domains of applications and related standardization needs currently envisaged by the CEN-CENELEC FGQT are: (a) novel applications enabled by QMSI devices; (b) characterization, benchmarking, and evaluation for reliable QT.

In analogy with the approach adopted in the Quantum Computing and simulation technologies, also in this case the following layers can be established to subdivide the ensemble of issues related to standardization concerning QMSI:
1. Quantum device(s)
2. Control electronics / optics / opto-mechanics
3. Control software

The first focus would be in the standardization of tools and technologies related to the quantum device, i.e. the quantum sensor. The central concept of a quantum sensor is that a probe interacts with an appropriate system, which changes the state of the probe. This effect known as quantum back action is intrinsically related to the physical nature of the measurement process: in this way, measurements of the probe reveal the parameters that characterise the system.

In quantum-enhanced sensors, the probe is generally prepared in a particular non-classical state. The best classical sensors exhibit a precision that scales proportionally to the square root of the number of particles N in the probe (i.e., the Standard Quantum Limit, SQL) whereas the best quantum sensors can in principle attain a precision that scales as N (i.e., the Heisenberg limit).

From the point of view of the quantum device, a useful classification [15] of the different domains and applications for implementing QMSI protocols (over various QT platforms, diversified from atomic systems to solid state and photonic devices) is schematized in the following:
- **Quantum Electronics**: such as, single-electron sources (for the SI-realization of the ampere and single-electron quantum optics, realization of the "quantum metrological triangle"), Josephson junctions (for quantum voltage standard); quantum Hall effect (for quantum resistance standard)

- **Quantum Clocks**: such as, ultra-cold atoms used for optical atomic or lattice clocks atomic clocks (for new time standards and chip scale atomic clocks to address Global Navigation Satellite System resilience, network synchronization, Time Stamping, Basic research, etc.)
- **Atomic Sensors**: such as, atom interferometers for gravimeter (for climate research, civil engineering, hydrocarbon and mineral exploration, GNSS-free navigation), magnetometers based on cold atoms or NV-centers in nanodiamonds (far brain imaging, hearth imaging, metrology, navigation); atomic vapour cells (for high precision electric magnetic and RF measurements)
- **Quantum Photonics**: such as, single-photon sources and detectors (for ultraprecise quantum interferometers, phase discrimination for quantum communication, twin beam and squeezed light, super-resolution, Sub-shot-noise imaging, quantum enhanced microscopy, quantum enhanced displacement sensing, quantum illumination and quantum radar/LiDAR, quantum reading, quantum ghost imaging and spectroscopy, quantum photometry and quantum physics based primary standards).

Where each individual QT platform can be cross-sectional with respect to the various domains, e.g. NV-centers platform is adopted both for magnetometer atomic sensors and for single-photon sources in quantum photonics; ultracold-atom optical lattice clocks are at the same time genuine atom interferometers and also atomic sensors leading to improved time and frequency standards, etc. Moreover, it is worth to mention that the examples above include both SI measurement standards used for traceability and novel QT-based measurement devices.

The requirements for proceeding towards the standardization are mostly platform specific, for instance, in the field of NV-centers in diamond one can outline the following list:
- **Quantum device(s):** requirements are related to technique of synthesis of the base material (native defect concentration spatial distribution, lifetime, coherence, lattice orientation), doping technique, and resolution of the implantation (a major breakthrough being the achievement of deterministic implantation), photo-luminescence properties (photon flux, spectrum)
- **Control Electronics/Optics/Opto-mechanics:** requirements are related to the parameters for DC and RF signal generation and processing, microwave power and polarization, geometry of the antenna delivering the microwave, eventual excitation pulse sequences
- **Control Software:** requirements are related to the programming of the measurement protocols and sequences, data displaying and storage

Concerning with the domain related to the characterization, benchmarking, and evaluation for reliable QMSI devices, the Cen-CENELEC FGQT envisaged that a general aspect to be taken into account in QMSI field refers to the complexity of its devices, which are based on the interplay of components from different partners. Constructing complex quantum devices out of these components requires dependable characterization with a "reliable data sheet" for the component under test as a result. This can be thought of in terms of a supply chain, one typical example being a foundry on one hand and a system integrator on the other hand. Only with this reliable data sheet, the quality and interoperability of QT components can be guaranteed – a prerequisite for practical applications with commercial perspective. Currently, offers for such independent characterization and testing capabilities are being developed in several National and European Programmes [21], [22], NMIs [23] and EMN-Q [15]. Testing and characterization capabilities need to be harmonised between testbeds. To put all this on a basis, standardised measurement protocols are needed. Ideally, this development goes hand-in-hand with the development of measurement capabilities and metrology for QT components not yet existing in many cases.

Benchmarking these devices can be non-trivial since the choice of performance indicators is not obvious and/or incompatible between different architectures. A set of standardised parameters and figures of merit allows benchmarking and comparing different approaches with respect to the application at hand. Quantum sensing devices can then, for instance, be measured against physical standards. Furthermore, standardised measurement protocols facilitate the comparison of quantum sensing/imaging against classical devices, crucial to evaluate perspectives of the technology. Advancing the capabilities above further facilitates future certification of QT components.

In conclusion, from an organisational point of view, given the general and cross-sectional character of QMSI field, the CEN-CENELEC FGQT emphasizes the need for a rationalization of efforts, to ensure that documentary standards that have been already prepared by existing SDO WGs/TCs for first generation Quantum Technologies (e.g. stable lasers, cryogenics, fibre-technology, chip-scale optical frequency comb, high-speed phase-coherent control of RF/microwave fields, integrated circuits, micro- and nano-fabrication capabilities for diverse materials, etc.) should be considered and upgraded taking into account Quantum-enabled technologies. In parallel, novel Technical Committees dedicated to Quantum Technologies should be created to oversee the standardisation process for the 'native' second generation QT (e.g. miniature atomic clocks, chip-scale magnetometers, microwave field imaging, atomic gyroscopes and Rydberg gas sensor, waveguide photonics, etc.) [3], [15].

## 3.4 Quantum communication

A general feature of all classes of Quantum Communication is the utilisation of a quantum channel for communication purposes. A quantum communication channel enables the spatial distribution of quantum states with high fidelity over an optical fibre, or with the help of optical telescopes through free space. Quantum communication can serve two purposes: either to perform more optimally a certain "classical" information communication task using a quantum channel, or to enable the transfer of quantum mechanical states over a distance with the goal of some subsequent utilisation in quantum computing or quantum metrology. While the first of these tasks in principle allows multiple applications, it has turned out that the only realistically efficient application to date is the simultaneous generation of cryptographic secrets at both communication ends. In contrast to the usual methods in conventional cryptography, these secrets (keys) remain secure against potential attackers independently of the attackers' resources. This cryptographic primitive is well known as Quantum Key Distribution (QKD). In this sense, both seemingly different Quantum Communication tasks are nevertheless based on the distribution of specific resources over a distance using quantum channels. Both tasks build on the transmission of quantum states, while for QKD these states have an auxiliary nature and it is the characteristics of measurements of these states (distant correlations), which are used to create the main resource, the cryptographic key, while the states themselves are being discarded. In contrast, in the second case, the quantum states themselves are the resource that is relayed between two communicating parties, either directly or indirectly. It needs to be underlined that the term "quantum communication" might sound misleading, as if it were a replacement of classical digital communication. Quite the opposite is true: essentially quantum communication needs always to be supported by classical digital communication in order to enable the distribution of additional quantum-enabled resources.

**Quantum Key Distribution**

Quantum Key Distribution is a general method ([24], [25], [26]) entailing two steps:
1. Utilisation of a quantum channel: distribution of quantum states end-to-end between two parties and ensuing distant measurements, yielding non-classical correlations in measured random data;
2. Classical post-processing: subsequent iterative communication rounds between the distant parties, involving local processing of the measurement data to distil identical random bit strings, completely unknown to external third parties. These strings can then be used as cryptographic keys in conventional cryptographic applications.

There is a huge diversity in QKD systems, as realised and even industrially produced already today (several of these can be seen in the equipment rack in Figure 4). With QKD, there are two main challenges: to ensure that the channel is really quantum (rather than some potentially flawed approximation), and to be able to perform the classical post-processing stage faithfully and efficiently. Often, advanced protocols utilise very clever combinations of both stages to ensure the final objective–an efficient generation of an end-to-end secure key.

Traditionally, QKD systems have been using a point-to-point communication medium (e.g. an optical fibre) to transfer quantum states (pulses of light). If we speak of light, ensuring the quantum nature of the signal means that very weak pulses need to be used. Light is unavoidably absorbed after some attenuation, i.e. communication distance, and this means that the range is restricted to typically a few hundreds of kilometres at best. The distribution of cryptographic keys over longer distances can only be achieved by means of composite systems, connecting several QKD links by means of "secure locations" (or trusted nodes) as interconnection points. These composite infrastructures are known as QKD networks of the first generation [27], [28].

Technologically still in the future, but otherwise conceptually realistic, is QKD over long distances by creating indirect quantum channels. This requires long-distance end-to-end quantum teleportation supported by end-to-end entanglement distribution relying on so-called quantum repeaters. These composite systems are the prospective Quantum Communication Networks of the second generation.

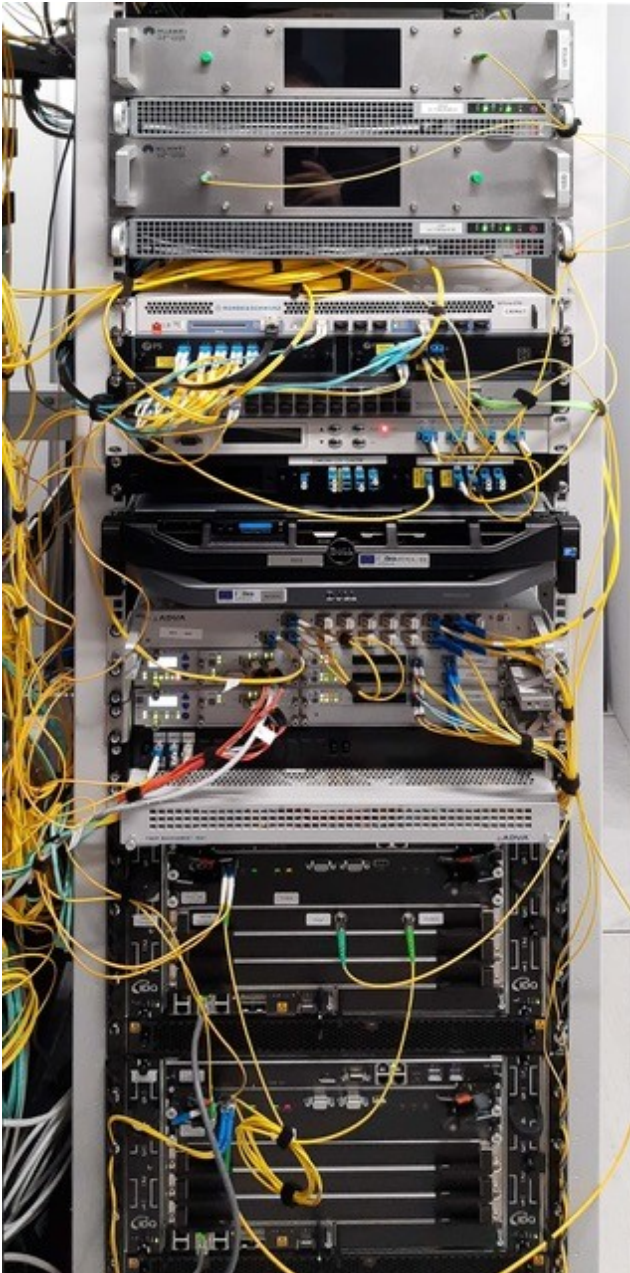

Figure 4: One of the central nodes in the Madrid QKD testbed, showing several QKD systems by different vendors in an equipment rack (Courtesy Vicente Martín, UPM Madrid)

**Quantum State end-to-end Distribution**

At first glance, it appears that this is a task identical to the first step of the QKD general method. Unfortunately this is an oversimplification, as even if a quantum state is successfully distributed to a receiving party, there is presently no quantum mechanism for this state to be "taken over" (actually received in the direct sense) by this party. The existing approaches rely on the mentioned quantum teleportation, in which the receiving party manages to create a state, identical to the one sent, at the price of destroying the first one and thus complying with the quantum no-cloning theorem. Similarly, as discussed above, an indirect quantum channel needs to be created, and only then, the input state can be teleported from the sender to the receiver.

Technically this is identical to the creation of an end-to-end quantum communication channel, as outlined above in the description of a Quantum Communication Networks of the second generation. For such networks, the second step of the general QKD method, namely the measurement of the "transmitted" state and the subsequent post-processing, is not needed.

It is, however, not actually reasonable to distinguish the design of (long-distance) end-to-end quantum communication channels for QKD or "just" for quantum state transfer: they are fundamentally identical. In this sense, all expressions of Quantum Communication Networks can be subsumed as Second Generation Networks (also known as Quantum Information Networks, or more expressively, although also a bit misleadingly, as the "Quantum Internet" [29]).

From the point of view of Quantum Technology, it is the development of these networks involving in turn the development of quantum repeaters, entanglement purification and teleportation protocols, which still represent future challenges of major importance.

It was actually in the field of Quantum Communication, where the standardisation of quantum technologies started in 2008 with the foundation of the ETSI Industry Specification Group for QKD (ETSI ISG-QKD), when the need arose for respective standards in a European funded research project. In the FP6 Integrated Project SECOQC, a first attempt toward security certification of QKD technologies was undertaken, which identified missing standards as a major roadblock. Since then, the group has produced several base standards, standards for interfaces, for components of QKD systems, for security proofs and security certification, and in other areas usually subject to standardisation. For about ten years, the ETSI was the only standards developing organisation active in this field—until other SDOs began to join in. Today, QT standardisation has become a major issue, with many SDOs active in the field, among them ISO/IEC, ITU-T, CEN-CENELEC, IEEE, the IRTF, as well as industry consortia, like the QuIC, and several metrology institutes. See Annex A for a summary of ongoing standardisation activities.

While several areas in quantum communication standardisation are already substantially covered, there are still gaps and blank spaces where much needed standards are missing.

This is especially true for the field of QKD security certification, where the ETSI ISG-QKD has produced a sample ISO/IEC 15408 "Common Criteria" protection profile for a prepare and measure QKD link, and the ISO/IEC JTC1 SC27 WG3 has produced two standards for requirements and test and evaluation methods for QKD products: Any security specification for a QKD product will need to rely on additional so-called background documents, especially where cryptographic protocols and algorithms need to be specified. All these cryptography-related choices will need to be based upon widely recognised and accepted background documents, like standards, or technical specifications. Evaluation laboratories will not accept the use of "proprietary crypto", not backed by respective background documents, i.e. standards. The specification of the quantum optical subsystem too will need to rely on compliance with external standards, e.g. for the employed QKD protocol, including its security proof, for specific components, like photon sources and detectors required by the QKD protocol, for random number generators, for attack methods and for attack rating methodologies. This issue has only arisen recently, and a timely identification of required background documents, i.e. standards, as well as the coordination of an efficient generation of these by standards developing organisations active in the field has become a major issue to prevent a potential major roadblock on the way to certified QKD products. Here the European CEN/CENELEC FGQT, as well as its envisioned CEN/CENELEC Joint Technical Committee successor, intends to assume a coordination function with the maintenance and dissemination of its QT standardisation roadmap.

After all, coordination of standardisation activities is becoming increasingly important in quantum communication (QKD) standardisation: It seems that most of the involved standards developing organisations are trying to cover the entire field of QKD standardisation by themselves, without looking much at the activities of other SDOs. This has currently already led to incompatible standards for security certification in the ETSI and ISO/IEC, as well as to several competing developments towards standards for cryptographic key management in QKD networks. If no effective steps are being taken to counteract that trend, the situation might end up as in the sector of IT cloud standardisation where dozens of standardisation developing organisations have developed virtually hundreds of competing standards, while the mostly used (de-facto) standards are set by major industrial players (Amazon AWS) outside the participative processes of actual standards developing organisations.

While in some fields multiple parallel approaches represent a potential "danger" and inefficiency, other areas, besides that of the mentioned background documents, still remain a blank slate: Currently there are no standards available e.g. for QKD protocols and for specific classes of photon sources and detectors. Networks for transferring quantum mechanical states ("full quantum networks") are even not at all covered in QT standardisation. Therefore, there is ample room for additional work to be carried out in quantum communication standardisation—with a substantial need for active coordination of activities to avoid the pitfalls of inefficiencies and double work.

## 3.5 Use cases and applications

With quantum technologies being a very broad field, so are the possible application areas. Developments in all areas of quantum technology bring these applications ever closer. However, even the best technology has little use if it is impractical in operational settings. It is therefore important that different aspects of quantum technology can properly interact and that it aligns with the classical technology already in place.

We identify two ways for a seamless interaction of new quantum technology with itself and with already existing technology. The first is one party developing everything themselves, which is reserved only for a small handful of parties. The far more common option is one party developing a part of a larger technology pipeline. The requires the input and output of the developed technology of one party to align with that of another party, as otherwise these cannot directly interoperate. Standards help interoperability as shown in the previous section.

This directly poses a challenge, as with the high-paced developments, interoperability and standards usually form a secondary priority and specifications can furthermore rapidly change. Additionally, if applications are still unclear, standard requirements might be hard to define.

Use cases can help identify future applications of quantum technology and different functionalities this technology should offer. Use cases also allow users to dream of ideal situations in which the technology is used and the accompanying functionalities.

Below we identify a few use cases and identify why they might help developing new future standards.
Historically, the first application of Secure Key distribution for enabling subsequent secure communication is the first use case of quantum technologies. From a practical point of view key distribution has to be performed without distance restrictions. To avoid the intrinsic limitation that absorption poses on quantum signals, the enabling quantum technology has been and will be augmented by the utilisation of composite quantum communication networks (of different generations as outlined below). First realisations are already near practical utilisation in telecom and government domains. For practical usage, it is also important how the generated secure key material is used in securing the future communication, as incorrect usage might leak (part of) the generated key.

Cloud-based quantum computing is one of the use cases where quantum computers are expected to prove useful. They will act as secondary processors hosted in the cloud and integration with classical high performance computers, allowing many computationally hard problems to be solved efficiently. This requires standard protocols for interaction between the quantum computer and classical devices. We also require standard methods to decompose workflows in smaller steps, each of which can be implemented on their respective device [31]. Finally, different components of the underlying quantum hardware have to interact, which requires standards if these components come from different vendors.

Possible applications for Quantum Metrology & Sensing include: (i) quantum magnetometers (based on cold atoms or Nitrogen-vacancy centres in nano-diamonds) for highly sensitive magnetic-field sensors, which have the potential to improve navigation and various imaging applications in arenas such as healthcare and Brain Imaging [30]; and (ii) Quantum-enhanced imaging, exploiting non-classical states of light to enable improved imaging performance, which is at the basis of ghost imaging (improving signal-to-noise ratio for applications at few photons level), quantum multiphoton microscopy, quantum coherence tomography, quantum interferometry and quantum lithography [16]. Standards for these applications include metrics to quantify the quality of the devices and calibration routines.

## 3.6 Supply-chain model for quantum technology standards

This section provides an abstract supply-chain model for QT standards. The purpose of this section is to provide inspiration for the initiation and development of standardisation for quantum-technologies.

The markets for quantum technologies are mostly still embryonic, and it is not known what these different markets or their full supply chains may look like. However, some markets are already emerging. In particular, there are emerging markets of scientists and technology developers who are building quantum platforms and systems from components that implement enabling technologies. Associated are markets for measurement equipment and measurement methods.

Let us start with an example from section 3.2 on quantum computing.

> EXAMPLE SUPPLY CHAIN: cabling of quantum computers
> Cryogenic quantum computers (e.g. transmon type) require a lot of cabling to control and read out the potentially thousands qubits. Whereas cabling and connectors are existing products, this extreme environment and specific applicant poses requirements that are different from regular (e.g. computer) cabling. Transmission requirements are different (e.g. on attenuation of control signals, and possible embedded passive or active functions), there are additional requirements on signal density, thermal isolation, vacuum insulation, outgassing and likely even more.
>
> As a consequence, this cabling ("control highway") is a specialised product with its specialised vendors, and it has a dedicated supply chain to purchasers that are developing and integrating cryogenic quantum computers.
>
> Standards will help both vendors and purchasers to characterise the relevant parameters of this product. Vendors can use the standard for marketing through benchmarking of their product against competitors. Purchasers refer to the standard to reduce the risk of a mispurchase.

This example can be abstracted into a more general supply-chain model for quantum technologies, see Figure 5. This model includes two or more counterparties ("Party A", "Party B") that engage in a transaction, e.g. a purchase. The transaction involves a product or service. Several aspects of that product or service may be standardised.

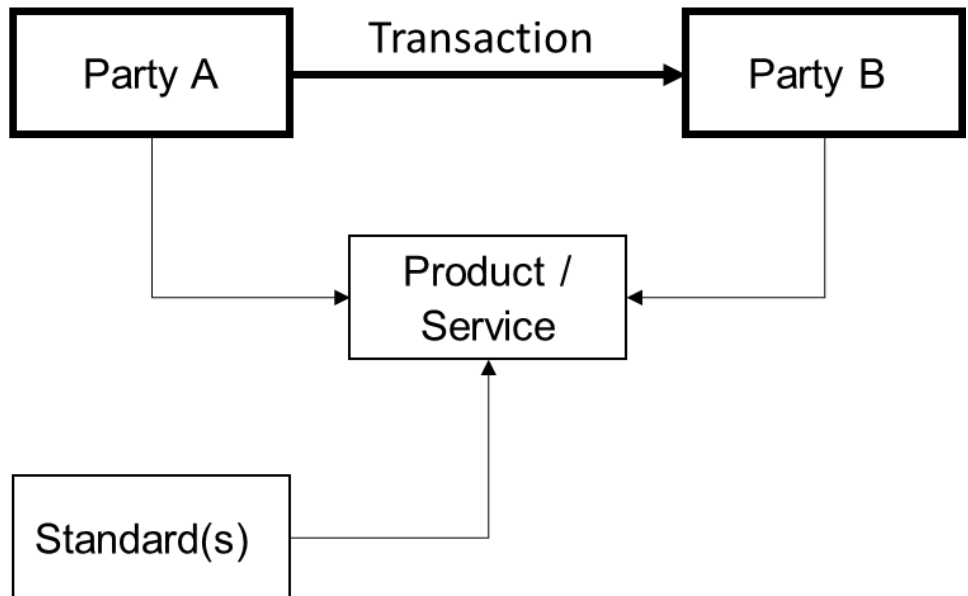

Figure 5. Abstract supply-chain model: a business transaction related to standards.

The following are further examples, albeit a bit more abstract, of where standards could be relevant in supply chains for quantum technologies.

**Example 1: Purchasing an ion trap**
Ion traps are a versatile quantum-technology component that can be applied to a variety of platforms and systems. A hypothesis is that there will be a market, where ion traps can be purchased as a unit that could be integrated into a platform or system. Standards may help enhance this emerging market for ion traps. One standardisable aspect is the characterisation of ion traps that enable comparing and benchmarking between different ion-trap products and vendors. Another standardisable aspect may be the connection interfaces to integrate an ion trap into a platform or system.

The product in this example is an ion-trap product. The counterparties are the vendors and purchasers of these ion traps. Standardisable aspects are technical characterisation and interface specification of ion traps. A benefit of standardisation would be the creation and growth of a market for ion-trap products.

NOTE: This example may be applicable to many more enabling technologies that could be productised.

**Example 2: Configuring a quantum simulator**
Quantum simulators enable the simulation of e.g. chemical processes, for the study of electronic and magnetic properties of particular systems or even design of new materials. A quantum simulator needs to be configured for a specific simulation task. Standardisation of the hardware components of a quantum simulator may provide information on its technical performance (e.g. quantifying with which fidelity the simulator can prepare an initial state and reproduce a specific

Hamiltonian). On a more generalised level, non-system-dependent standards may provide a language to configure a quantum simulator.

The product in this example is a configurable quantum simulator. The counterparties are a chemical engineer who orders a quantum simulation, and a quantum engineer who configures/constructs the requested quantum simulator. A standardisable aspect is the language that describes the configuration. A benefit of standardisation would be the creation and growth of a market for quantum simulators.

**Example 3: Developing an algorithm for a cloud quantum computer**
Similar to the previous example would be quantum-computing capacity that is offered as a cloud service by multiple competing cloud-quantum-computing service providers.
The service in this example is a cloud-computing service. The counterparties are a developer with a quantum-computing task and a cloud-quantum-computing service provider. A standardisable aspect is an application programming interface. A benefit of standardisation would be the creation and growth of a market for cloud-quantum-computing services.

**Example 4: Installing a QKD ground station**
Quantum Key Distribution (QKD) enables a highly secure generation and distribution of cryptographic keys that does not rely on the unbreakability of mathematical algorithms. QKD has already been demonstrated in the space domain, involving orbital satellites. It is envisioned that a QKD network could be constructed with multiple satellites, multiple ground stations and multiple operators that operate different parts of the QKD network.

The product in this example would be the installation of a new QKD ground station. The counterparties are the operator of the ground station and the equipment vendor that provides the QKD equipment for the ground station. A benefit of standardisation would be the creation and growth of a market for QKD equipment, as well as avoidance of vendor lock-in towards a specific QKD technology provider.

**Example 5: Purchasing quantum-technology measurement equipment**
The specifications of quantum-technology platforms and systems, as well as their enabling technologies need to be verifiable. Measurement equipment is required for this. Standards would specify the measurable aspect, and what are the technical requirements that a specific measurement should satisfy.

The product in this example would be measurement equipment that is used in a quantum-technology context. The counterparties are the purchaser and vendor of the measurement equipment. Standardisable aspects are the measurement method itself, as well as characterisations of the fidelity of the measurements. A benefit of standardisation would be the creation and growth of a market for measurement equipment for quantum technology set-ups.

**Example 6: Submitting a scientific article for peer review**

Related to the previous example are scientific articles that involve measurements. If a measurement method is specified in a standard, then the author can reference that standard, instead of completely detailing it.

The product in this example is the submitted scientific article. The counterparties are the author and the peer reviewer. The standardisable aspect would be the applied measurement method. A benefit of this standardisation is that it would make scientific articles on a specific measured feature better comparable, enabling benchmarking the results from the scientific article against its cited references.

## 3.7 Structuring standardisation of quantum technologies

This section concludes with an attempt to converge the storyline of the FGQT roadmap, and to provide some structuring to the various aspects of quantum technologies.

Quantum Technologies are commonly structured in domains. The European Quantum Flagship roadmap [6], [17], [18], for instance, features four domains or 'pillars' covering the whole range of QT. For the identification of standardisation needs, however, this classification brings the disadvantage that there are strong commonalities between those pillars, as for example some enabling quantum technologies that are equally applicable to all domains with associated parallels in their standardisation perspective. Within discussions in FGQT, this challenge was tackled and an overarching framework for the identification of standardisation needs was developed that underlies the work-in-progress roadmap. The remainder of this section highlights the view on this framework developed by FGQT during 2021.

A common way to categorise the range of QT is to think in terms of pillars, namely quantum communication, quantum computing & simulation and quantum metrology, sensing & enhanced imaging. This is, however, an oversimplification due to many matrix-like connections as we discuss below.

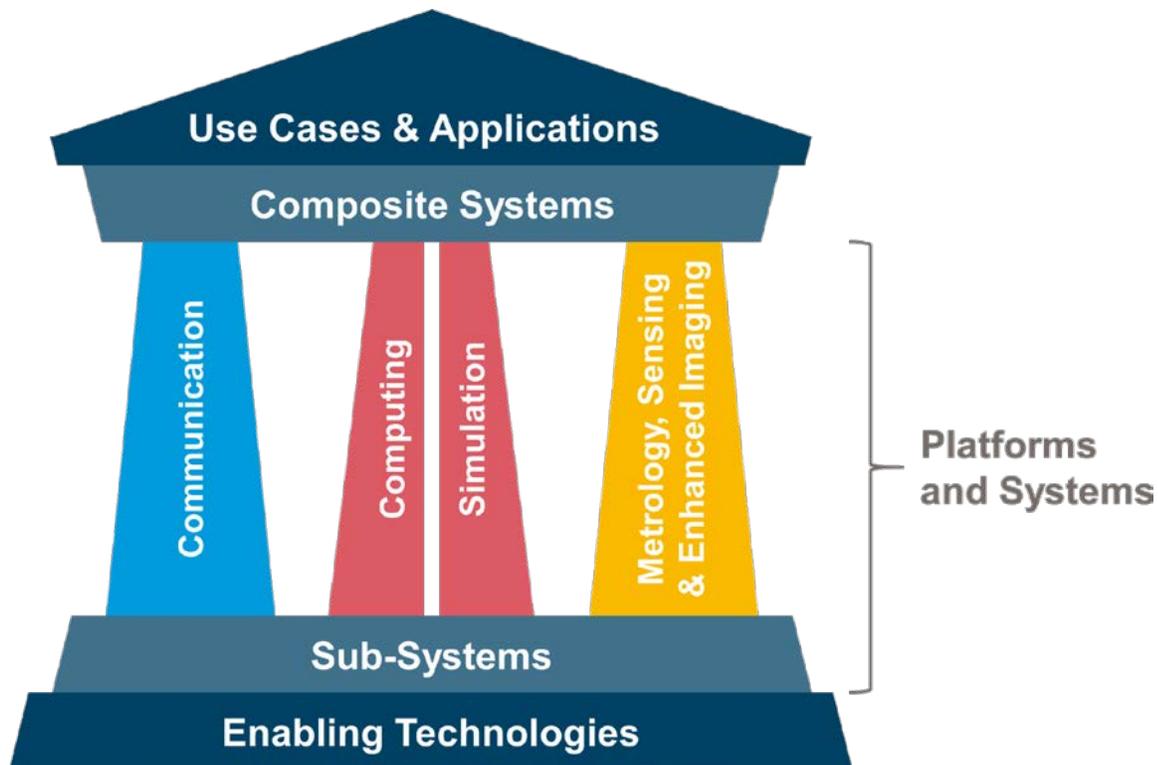

Figure 6: Quantum technology "temple" structure (Source: [32])

Indeed, this is common for all QT underlying hardware or "enabling technology" that facilitates the design and prospective manufacturing in the QT pillars/domains mentioned above. Accordingly, the tools, as for instance software, used for controlling quantum states are typically universal. Combining elements from enabling technologies and tools facilitates the assembly of subsystems, which can be combined to create QT platforms, systems and higher-level composite systems or infrastructures. These various technological levels then give rise to societally relevant applications to be grouped in general use cases.

We propose to consider these horizontal layers common to several or all fields of QT, representing different levels of technological complexity and proximity to applications and use cases jointly with the traditional pillars. Naturally the latter are still highly relevant, but their connections to each individual horizontal layer, representing a "matrix structure", need to be included. In Figure 6, we illustrate this idea of a hierarchy or matrix of QT with the architecture of a Greek temple. In this structure, standardisation needs can naturally be identified by connections between the different horizontal and vertical layers. For instance, interoperability between the different layers requires well defined interfaces. This idea is further reflected in Figure 7, where we explicitly show some connections, which are natural points of identification of standardisation needs. Working in this picture that includes all fields of QT requires a group of experts with the corresponding breadth of expertise. We note that these connections strongly depend on the development of QT components and might not be obvious to identify, but the matrix structure helps to guide the development. Furthermore, the nature of the standardization need might also depend on the layers in Figure 7. For instance, whereas in the "enabling technology" layer, specifying characteristic properties (for example, material properties like NV center purity or density; or heating rate of an ion trap, or noise figure of a TWP amplifier) might be typical, in the "platform" layer a performance characteristic might play a significant role, as for instance, sensitivity of a sensing device.

There is currently a multitude of standardisation activities worldwide in quantum technology and some but not all sub-areas. In many cases, these activities are constrained to a certain sub-field and, correspondingly, to a specialised expertise. While this has the advantage of providing a relatively straight-forward means to arrive at specific standardisation actions, the larger picture of QT as sketched in this document cannot be addressed. The underlying matrix structure as discussed above cannot easily be represented. In particular, standardisation needs common to several of the sub-fields might be addressed in an incompatible way, potentially leading to roadblocks later on. Consider a component from the "enabling technologies" layer that is used in two QT domains/pillars. For example, a certain material might require standardized parameters to qualify for a quantum sensing application (could be a NV-center). The same material could be used for a quantum computing application, however, the relevant parameters are not necessarily identical. Another example could be a control software, or a device (laser) for manipulating a quantum state. In all those cases, one common standardized set of parameters (of course, with values ranges corresponding to the application at hand) constitutes a sufficient and complete description. At the same time an incoherent, incompatible or incomplete description, valid for one application but not for the other, is prevented. In the worst case, different and specialized standards are developed specifically and independently for each application and domain which might limit the usefulness and thus create a roadblock for effective standardization.

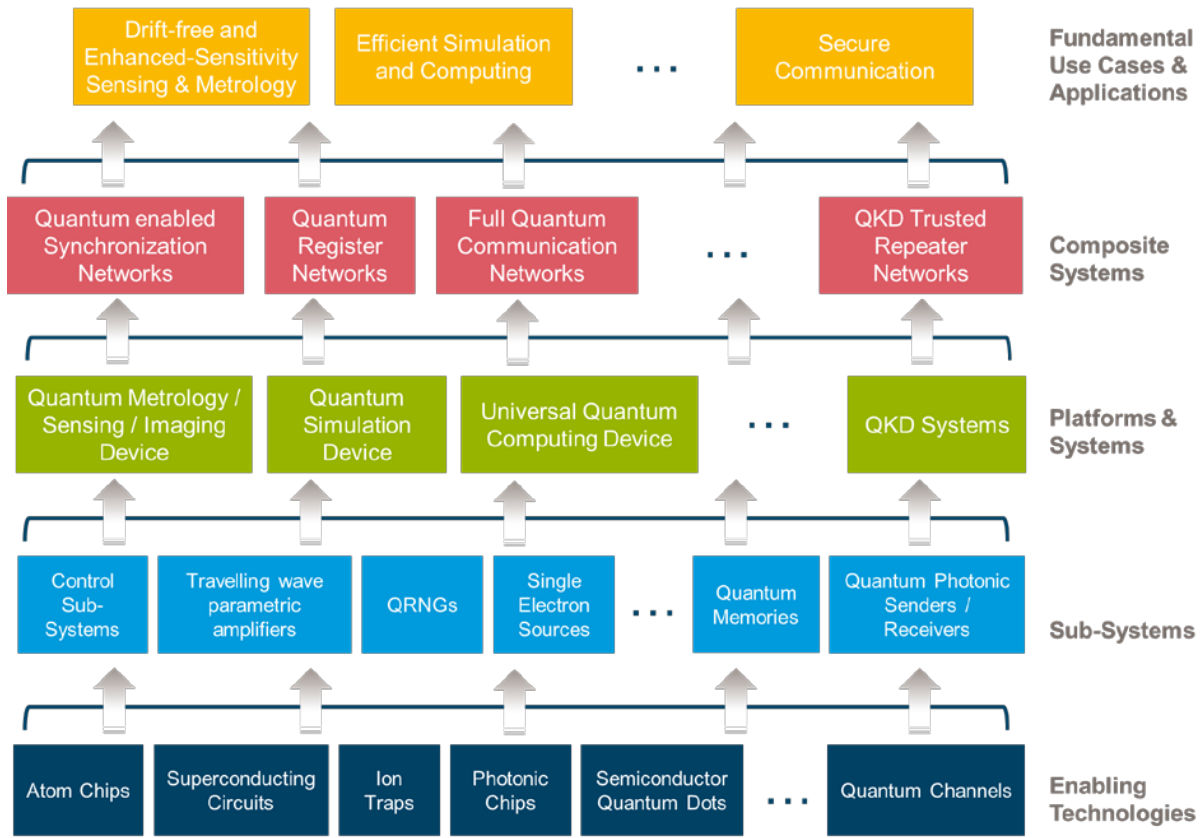

Figure 7: Connections between different technology levels (Source: [32])

# 4. Organisations perspectives

## 4.1 General

Organisations may have many reasons why to contribute to standardisation. Standardisation is a good place for networking with other organisations that have similar interests. Academic and research organisations may use standardisation to showcase their technologies and competence. Organisations that purchase technologies may use standardisation to coordinate use cases and requirements, making sure that upcoming products address their market needs, and less need for expensive proprietary solutions and their associated vendor lock-in risk. Service providers may use standardisation to coordinate with their suppliers, using the regulated SDO environment to assure fair competition. Suppliers of technologies may coordinate with competitors on non-competitive aspects of their product, to reduce market fragmentation and achieve critical mass for new product categories. Regulators may contribute to standardisation, to ensure that regulatory requirements can be technologically met. Other organisations may have other reasons, which include consultancy and patent licensing.

Section 4.2 provides some statistics on FGQT and its delegates. Section 4.3 summarises the rationale of the co-authors and their organisations to actively contribute to FGQT. Annex B provides more detailed insights from these organisations, why they have been contributing to FGQT, and what they would like to achieve.

## 4.2 FGQT statistics

When FGQT kick-off mid-2020, the delegates were asked about their affiliation from which they participate in FGQT. More than 50% of the delegates had an academic affiliation, 30% industry, and 11% standards, see Figure 8. This highlights the early stage that QT standardisation is still at, as standardisation for mature markets is dominated by delegates with an industrial affiliation, contributing to standards with a specific product or service interest.

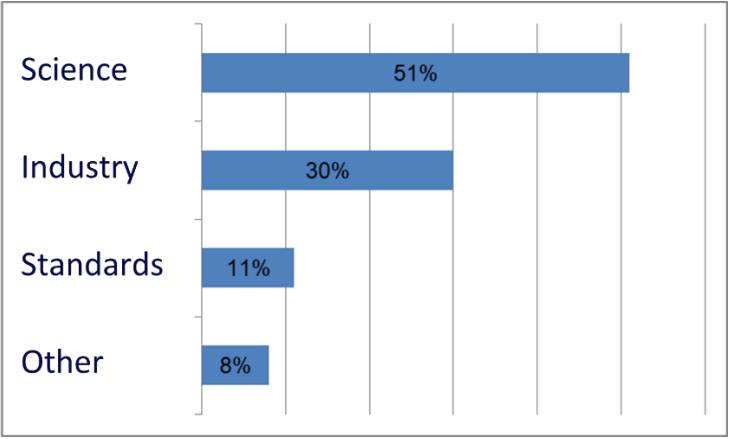

Figure 8: Background of FGQT delegates (mid 2020).

Most European countries are represented in FGQT, with some dominance of Germany, Italy, UK, Switzerland and Netherlands, See Figure 9.

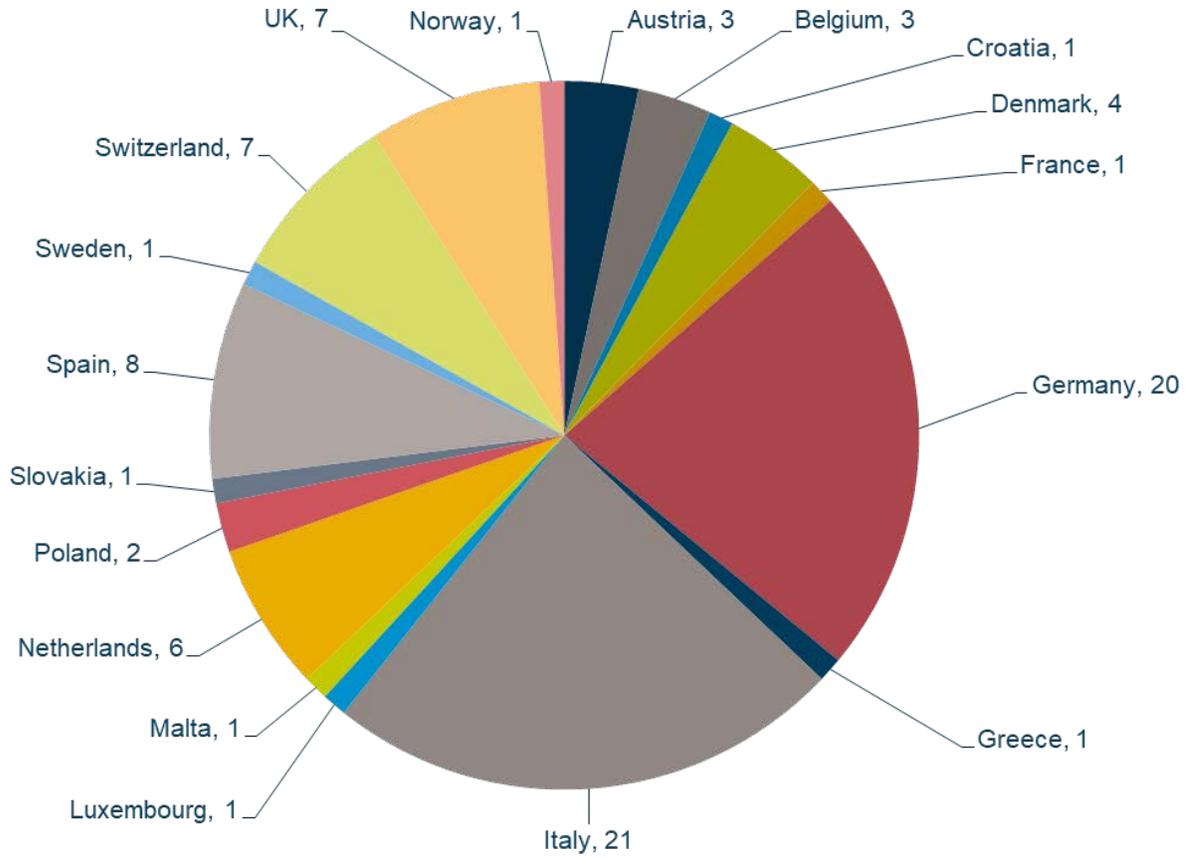

Figure 9: FGQT interest from the various European countries (mid 2020).

Mid-2022, the FGQT member list counts over 110 delegates. FGQT has had 27 meetings to this point, mostly via video and two physical meetings. There have been over 250 contributions to the FGQT roadmap and use-cases deliverables.

## 4.3 Rationale of FGQT delegates

Figure 10 summarises the main area of contribution of the most active FGQT contributors, the co-authors of this article. Annex B provides more detailed about each. Note that the organisations labelled as "academic/research" each contribute to one or more of the areas quantum computing, quantum communication and/or quantum metrology.

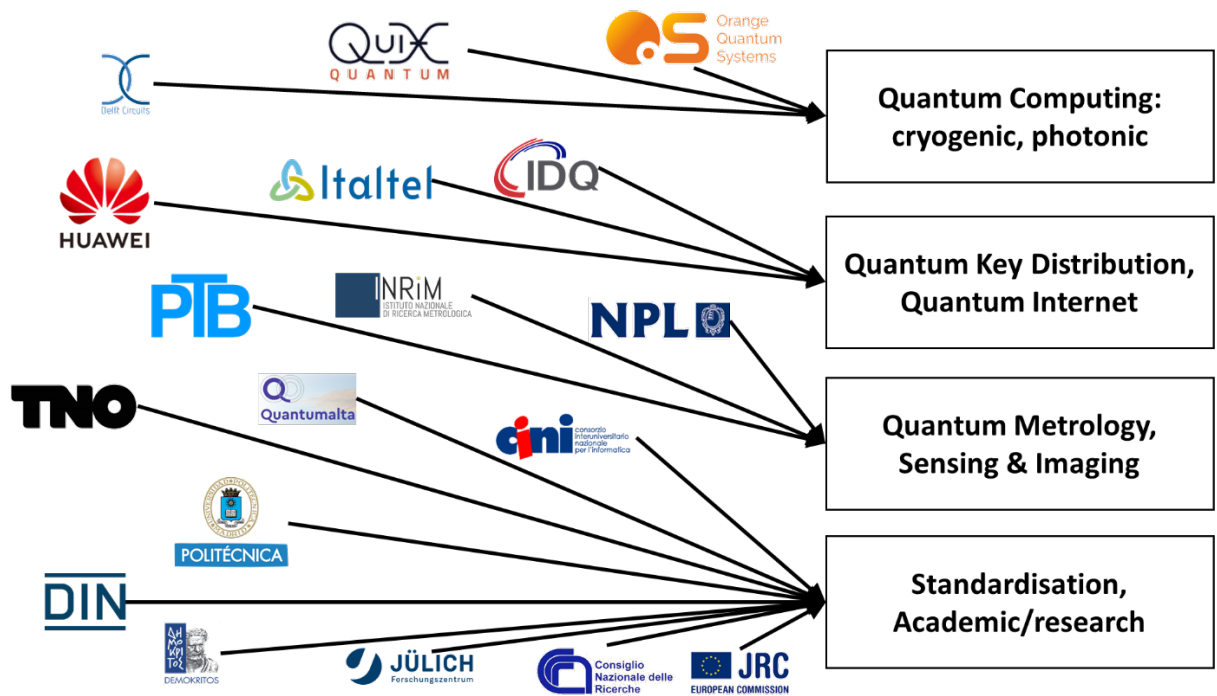

Figure 10: Active FGQT contributors (mid-2022).

# 5. Conclusion: work in progress

This review article makes clear that standardisation for quantum technologies is still at a very early stage. Most quantum technologies still have low technology readiness levels. Nevertheless, supply chains are already emerging for some quantum technologies. For example, research infrastructures for quantum computing are quite capital intensive, and standardisation may help rationalise and speed up this research. Commercial products are also becoming available in quantum sensing, metrology and imaging, where standardisation may help structuring and growing the market. Published standards on quantum key distribution are already available. On the other hand, entanglement-based interconnection between quantum computers only exists as a concept.

More analyses are needed to determine "standardisation readiness" of the different quantum technologies, and when the market is expected to be ready for formal SDO development of technical guidelines (best practises), technical reports (use cases, requirements), and technical specifications (interoperability). For this purpose, the CEN-CENELEC FGQT is developing a roadmap for standardisation relevant to quantum technologies. Its work includes coordination with other SDOs and industry forums, and it is expected to spin off actual standardisation work.

# Annex A. Related work

## A.1 General

This Annex provides further details abouts relevant quantum-technologies-related standardisation activities for section 2.

## A.2 ETSI: quantum key distribution

The ETSI Industry Specification Group on Quantum Key Distribution (ISG-QKD, [33]) was the first forum aiming at standardisation of quantum communication technologies. It was created in 2008 by a group of QKD manufacturers, telecommunications providers, universities and research institutions, mostly from Europe. Since then the group has grown with the incorporation of new members. Today it has representatives from the main actors in QKD in Europe and also from the USA, South Korea and Japan.

ISG-QKD has produced well over 20 documents. Although several of them (e.g. Vocabulary Group Specification GS QKD 002, [34], Use Cases GS QKD 007 [35]) had a general scope, the group has concentrated more on specific technical aspects than on framework documents. To some extent, this reflected the interest of its members, who are more concerned with the necessity to have standards supporting the creation of QKD products than with general paradigms of Quantum Technology. It also reflects the fact that when the ISG-QKD started, the technology was less mature and many technical aspects had to be covered before dealing with broader views. As such, fundamental interfaces, like those needed to extract a key from a QKD device (GS QKD 004, [36] and GS QKD 014, [37]) or to control devices (GS QKD 015, [38]) have been published. Fundamental aspects of QKD Security such as "QKD security implementation white paper" [39], GS QKD 005, [40], GS QKD 008 [41] have been addressed. A further field of interest has been the characterizations of QKD components or security proofs and specifications have also been addressed (GS QKD 003, [42], GS QKD 011, [43]). At present a number of the mentioned documents are being updated. Areas of intensive work are related to security certification of QKD (Work on Common Criteria Protection Profile in collaboration with BSI Germany) and the analysis of existing approaches on standardisation of QKD-Networks.

The current work continues with the definition of other interfaces (e.g. network orchestration GS QKD 018, and authentication GS QKD 019), that will be needed to create larger networks, and with the important task of security certification, where an ISO/EN 15408 "Common Criteria" Protection Profile [44] for a QKD link is being finished (GS QKD 016). Although this is likely the most advanced effort of any SDO towards QKD security certification, it is only the first step within the Common Criteria scheme, albeit a much needed one to have QKD fully accepted as a high security technology in a broad market.

## A.3 ITU-T: quantum information technology for networks

ITU's work has been concentrated mainly in the area of quantum communication and more generally has been limited to implications of quantum technologies on communication and communication networks. The work on QKD networks and security aspects of the latter is led by ITU-T Study Group 13 (Future networks and cloud [45]) and ITU-T Study Group 17 ("SG17 Security", [46]). Some work on quantum random number generation has also been carried out in SG17. The centrepiece has been the outline of ITU standards for QKD networks, including provision of foundational concepts (ITU Y.3800, [47]), address functional requirements (ITU Y.3801 [48]), architecture (ITU Y.3802, [49]), key management (ITU Y.3803, [50]), and control and management (ITU Y.3804, [51]). ITU standards also provide a security framework for QKD networks (ITU X.1710, [52]), key combination methods (ITU X.1714, [53]), and the architecture of a quantum noise random number generator (ITU X.1702, [54]).

These ITU standards for QKD networks aim at enabling the integration of QKD technology into large-scale ICT networks and provision of the security of the latter.

The ITU-T Focus Group on Quantum Information Technology for Networks (FG-QIT4N) [55] studied the evolution of quantum information technologies in view of their foreseen applications in ICT networks. The group was conducting exploratory 'pre-standardisation' studies to identify emerging standardisation demands and anticipate demands to arise in future.

The FG-QIT4N was established in September 2019 to provide a collaborative platform for interested stakeholders - such as researchers, engineers, practitioners, entrepreneurs and policy makers - to share knowledge, best practices and lessons learned to take full advantage of the ability and potential of QIT in networks. Its main objectives were:
- to study the evolution and applications of QIT for networks;
- to focus on terminology and use cases for QIT for networks;
- to provide necessary technical background information and collaborative conditions to effectively support QIN-related standardisation work in ITU-T Study Groups;
- to provide an open cooperation platform with ITU-T Study Groups and other SDOs.

The Focus Group had been organised as follows: two main research groups - one dedicated to Quantum Key Distribution Networks (QKDN), one dedicated to Quantum Information Networks (QIN) that are beyond QKDN - and a management group The term of the FGQIT4N expired in December 2021 and resulted in 9 deliverables / reports [56]. The management group continues to work on a summary of the outcomes.

## A.4 ISO/IEC JTC1: quantum technologies standardisation

The ISO and IEC in their Joint Technical Committee JTC1 have quantum technologies standardisation activities in two working groups (WG): WG14 on Quantum Computing, and Sub-Committee SC27 WG3 for security certification of QKD systems.

Working Group 14 on Quantum Computing [57] was established with the following objectives: to serve as a focus of and proponent for JTC1's standardisation program on Quantum Computing; to identify gaps and opportunities in Quantum Computing standardisation; to develop and maintain a list of existing Quantum Computing standards produced and standards development projects underway in ISO/TCs, IEC/TCs and JTC1. Another objective is to develop further deliverables in the area of Quantum Computing. As a systems integration entity, it maintains relationships with other ISO and IEC technical committees and with other organisations that are involved in Quantum Computing standardisation.

The first work item of WG14 is ISO/IEC-4879 [58] which is to develop a standard for terminology and vocabulary for quantum computing. Work started in 2020 and the committee draft is expected to be ready during the first half of 2022. A full standard should be ready towards the end of 2022. When ready, this will be the first standard specifically developed for quantum computing by a standards development organisation.

SC27 WG3 [59] is the working group that develops and maintains the ISO/EN 15408 "Common Criteria for Information Technology Security Evaluation" [60] and has two standards for the security evaluation and certification of QKD systems in a quite advanced "Committee Draft" stage (planned publication still in 2022): "ISO/IEC 23837-1 Information security—Security requirements, test and evaluation methods for quantum key distribution" Part 1 "Requirements" (containing predefined security functional requirements for use in QKD PPs) and Part 2 "Test and evaluation methods" [61].

## A.5 IEEE: emerging quantum information market

The U.S. Quantum Economic Development Consortium (QED-C) and international counterparts have expressed interest to IEEE in developing standards appropriate for the emerging quantum information market [62]. According to IEEE, quantum information standards are likely to evolve over time from informal efforts to formal specifications ([63]). A formal, international quantum standard starts when companies or individuals working in an area approach IEEE with a proposal called a Project Authorization Request (PAR). Currently, there are four active quantum standards projects, which are briefly described below.
- P1913 – This standard defines an application-layer protocol denoted as Software-Defined Quantum Communication (SDQC) that communicates over TCP/IP and enables configuration of quantum endpoints in a communication network to dynamically create, modify, or remove quantum protocols or applications. Moreover, SDQC includes a set of commands that control the transmission, reception, and operation (i.e., preparation, measurement, and readout) of quantum states. [64]
- P7130 – This standard is related to specific terminology for quantum technologies, establishing definitions necessary to facilitate clarity and understanding to enable interoperability and compatibility. [65]
- P7131 – This standard covers quantum computing performance metrics, with the objective to standardise performance benchmarking of quantum computing hardware and software. The considered metrics and performance tests enable the evaluation of quantum computers standalone or by comparison against quantum and classical computers. [66]
- P3120 – This standard defines technical architectures for quantum computers, including hardware components and low-level software (e.g., quantum error correction). [67]

## A.6 IRTF: quantum internet research

The Internet Research Task Force (IRTF) has hosted the Quantum Internet Research Group (QIRG) [68] since the IETF 101 meeting in March 2018. The QIRG has no official membership and participation is open to everybody. The Research Group communicates primarily through its mailing list which can be freely subscribed and posted to [69]. The entire mailing list archive is publicly available online [70]. The QIRG also holds two or three meetings per year, virtually or in-person, usually at the IETF meetings.

The scope of the QIRG's work is defined in its charter [68]. A key goal of the QIRG is the development of an architectural framework delineating network node roles and definitions that will serve as the first step toward a quantum network architecture. However, it is important to note that the QIRG focuses on fully entanglement-based quantum networks. QKD and trusted repeater networks are also often discussed, but usually in the context of being a stepping stone towards such a full quantum internet.

The QIRG, just like all the other IRTF Research Groups, does not work on standards. It is instead focused on developing research collaborations and teamwork in exploring research issues related to the Internet. Nevertheless, the Research Group does also work on producing technical documents on quantum networks [70]. Currently, it is working on two Internet Drafts which the group aims to publish as informational RFCs (i.e. not standards specifications):
- Architectural Principles for a Quantum Internet [71],
- Application Scenarios for the Quantum Internet [72].

Since quantum networks are so different when compared to classical networking, the QIRG is also focused on educating the classical networking community on this new subject. In addition to discussions on the mailing list, the QIRG also hosts seminars with speakers from both industry and academia. So far three such seminars have taken place:
- "Practical Quantum Networking at Room Temperature" by Mehdi Namazi (Qunnect Inc.) [73],
- "Genuine and Optimized Entanglement-Based Quantum Networks" by Wolfgang Dür (University of Innsbruck) [74],
- "Building Quantum Networks at the Local-Area Scale" by Marc Kaplan (VeriQloud) [75].

## A.7 EURAMET EMN-Q: quantum metrology coordination

A group of European National Metrology Institutes (NMIs) have recently created a European Metrology Network for Quantum Technologies (EMN-Q) [15] under the auspices of EURAMET [76] to tackle this technological paradigm shift. Large companies, as well as start-ups, have started to develop and engineer quantum devices or begun to integrate them into their products: the commercial success of QT, together with progress in research and development, relies on certification and reliability built upon internationally agreed standards and metrological traceability.

Therefore, the objective of the EMN-Q is to coordinate the activities of the European NMIs to ensure their efficient support for European competitiveness in quantum technologies. A special focus of the EMN-Q will be to develop new measurement capabilities and dedicated services to serve the rapidly-growing needs of industry and research institutions in this field.

Industry, governmental agencies, academic sectors or any other type of stakeholder are welcome to contact the EMN-Q and discuss their metrology needs. These can relate not only to quantum characteristics of quantum devices, but also to metrology of key enabling technologies, metrology that can improve the supply chain of industrial quantum devices or other industrial needs connected with quantum technologies.

The commitment of the EMN-Q is to become the unique contact point to stakeholders interested in metrology for quantum technologies by:
- contributing to standardisation & certification of quantum technologies;
- promoting the take-up of metrology in the development of these technologies;
- supporting industrial needs in synergy with the technological objectives of the EC Quantum Flagship and national quantum technology programs;
- promoting the use of quantum measurement techniques where advantageous for "classical" technical areas.

The EMN-Q is developing Roadmaps and a Strategic Research Agenda to identify priorities for research by Europe's national metrology institutes and designated institutes and to identify collaboration partners for such research.

The European Metrology Network (EMN) for Quantum Technologies will support the integration of measurement science with quantum technologies in three sections: Quantum Clocks and Atomic Sensors, Quantum Electronics and Quantum Photonics [23].

## A.8 QuIC: Industrial coordination for standardisation

The European Quantum Industry Consortium (QuIC, [77]) is a European not-for-profit business association aiming to build a strong, vibrant ecosystem between business actors and leading research and technology organisations in the Quantum Technology domain. QuIC believes that only a strong and unified quantum technology community in Europe will be able to succeed in the current global race to become the center of the next technological revolution. QuIC organises the work toward its objectives in Working Groups.

As QuIC is the voice of quantum industry in Europe, the Standards Working Group (WG4) aims to become the unified voice of quantum industry in Europe on standardization issues. It will provide a single point of contact to voice the needs for standardization from the industry to decision makers, politics, and standardization bodies.

The Standards Working Group plans the following actions.
- Foster communication between the QuIC members and the Standardisation bodies and facilitate the creation and interchange of information. Within this realm, WG4 will organise communication events with SDOs and other relevant Groups to build awareness and promote standardisation activities for Quantum technologies among the members of the QuIC;
- Set up a methodology and the tools for eliciting the standardisation needs coming from the Industrial members of the QuIC and communicate these needs with the SDOs.
- Develop a living document "State-of-the-art on standardisation activities in Quantum Technologies". The document will present among the others the updated information on the activities of SDOs and the upcoming standards;
- Create twinning activities with the other Working Groups of QuIC
- Support QuIC in running projects such as QUCATS.

The QuIC Standardisation Working Group will develop standards itself, since its role will be supportive of SDOs activities.

## A.9 StandICT.eu: EU funding for ICT standardisation

The StandICT.eu 2023 project is a Coordination and Support Action of the EU Horizon 2020 framework programme. It started in Sept. 2020 and has as its "(...) central goal to ensure a neutral, reputable, pragmatic and fair approach to support European and Associated states presence in the international ICT standardisation scene." [78] To this goal, the project issues ten open calls with funding opportunities for European experts in several strategic fields, including the field of Quantum Technologies. Currently, the StandICT project supports an editor of the FGQT's Quantum Technologies Standardisation Roadmap.

# Annex B. Organisations perspectives

## B.1 General

This Annex provides further insights from the co-author's organisations, why they have been contributing to FGQT, and what they would like to achieve. A summary overview is provided in section 4.

## B.2 TNO: quantum technologies & standards

TNO, The Netherlands Organisation for Applied Scientific Research [79], is the largest Dutch national institute for applied research. Its research portfolio includes a wide spectrum of quantum technologies [80], as well as their applications and quantum-safe cryptography. As the mission of TNO is to bridge the innovation gap between technology and market, TNO has always invested heavily in standardisation activities as a great place to develop and coordinate technology knowledge, to be applied in joint projects with European partners, Dutch industry and government.

TNO is one of the co-founders of FGQT and provides its first chair position, partially financed by the European funding from the QFlag-CSA [81], QUCATS-CSA [82], Dutch funding from QDNL [83], as well as own TNO-internal investment. TNO was also one of the co-founders of the Dutch NEN-FGQT "mirror group" where Dutch FGQT contributions are coordinated. Specific quantum-technology standardisation areas for TNO are the concept of quantum cloud computing, and the coordination with subsystem vendors for cryogenic quantum computers for the Delft QuTech collaboration.

## B.3 PTB: quantum metrology

The Physikalisch-Technische Bundesanstalt (PTB) [84] is the National Metrology Institute of Germany and is responsible for the realisation and dissemination of the SI units, as well as basic metrology research for society and for industry. Pushing precision metrology to its limits usually requires quantum technology. PTB offers extensive resources and infrastructure for a large scope of precision measurements.

In 2019 the PTB founded the Quantum Technology Competence Centre (QTZ), which supports industrial developments in the field of quantum technologies. The QTZ operates dedicated user-facilities for the testing of emerging quantum technologies, including, for instance, quantum electronic devices, optical components and ion traps employed for atomic clocks and quantum computing. PTB is developing independent characterisation capabilities for QT components, ensuring quality and reliability. This goes hand-in-hand with pushing forward standardisation of QT. PTB provides the vice-chair of FGQT and member of the European Metrology Network for Quantum Technologies (EMN-Q).

## B.4 DIN: standards for Germany

DIN, the German Institute for Standardisation [85], is the independent platform for standardisation in Germany and worldwide. It was founded in 1917 and celebrated its 100 year anniversary in 2017. As a partner for industry, research and society as a whole, DIN plays a major role in helping innovations to reach the market in areas such as the digital economy or society, often within the framework of research projects.

More than 36,000 experts from industry, research, consumer protection and the public sector bring their expertise to work on standardisation projects managed by DIN. The result of these efforts are market-oriented standards and specifications that promote global trade, and encourage rationalisation, quality assurance, and the protection of society and the environment, as well as improving security and communication.

DIN represents German interests in international organisations such as CEN, the European standards body, and ISO, the International Standards Organization. Today, roughly 85 % of all national standards projects are European or international in origin. International standards provide a common language for the technical world, supporting global trade.

Particularly when it comes to innovative topics, such as quantum technologies, DIN tries to identify standardisation potential at an early stage and helps the experts to develop the necessary standards. In the field of quantum technologies, DIN holds the secretariat of the CEN-CENELEC *Focus Group on Quantum Technologies* (FGQT) and the ISO/IEC JTC 1/SC 27 *Information security, cybersecurity and privacy protection*.

## B.5 Huawei Technologies Duesseldorf GmbH: advanced CV-QKD prototypes

The Quantum Communication group at the Munich Research Center (MRC) of Huawei Technologies Duesseldorf GmbH [86] was initiated in the fall of 2015. It is part of the Optical research in MRC. The goal from the very beginning has been to develop practical Continuous Variables (CV) QKD technology. The applied research team has changed over time but has comprised experts in theory and design of QKD, signal processing, as well as software and FPGA design and development.

The goal has never been to implement laboratory systems that are suited for setting specific performance records but rather wholesale prototypes suited for autonomous field operation that are simultaneously software-defined and adapted to seamless steering and control. This program has been successful and already in the spring of 2018 three prototype devices (one sender and two receivers) has been prepared and shipped together with OSN 1800 Huawei devices equipped with encryption cards for an initial test at the premises of Telefonica in Madrid (a dedicated network but connecting real life central offices with no direct access of the team to the devices). A revolutionary innovation has been the ability of the sender to directionally switch to one of the two receivers on demand.

The team participates in the EC Quantum Flagship project CiViQ and later became a member of the Industry Board of the European Open QKD project. In this capacity the team delivered in the beginning of 2021 10 QKD devices to the OpenQKD Test Bed in Madrid. This is a new generation that features both directional and wavelength switching on demand, as well as dual optical polarisation and rapid FPGA-powered pre-processing, The 5 senders and 5 receivers are situated in 7 locations across two operator networks in Madrid, establish up to 25 links over significant distances and interoperate with optical equipment, encryptors and QKD devices of other providers.

## B.6 Delft Circuits: hardware for quantum engineers

Delft Circuits (founded in 2017, Delft, Netherlands, [87]) is a rapidly growing company, with a focus on creating hardware solutions for dense i/o chains in quantum computers. Creating the control highway for quantum computers with 1000 qubits or more requires another approach than just trying to put thousands of coaxial cables in a single cryogenic fridge, with many interconnections at each thermal stage.

The present showpiece is a scalable cryogenic multi-channel i/o solution called Cri/oFlex®, suitable from DC to microwave frequencies. This solution combines low-thermal conductance with excellent thermalisation capabilities and (super)conducting circuits on flexible substrates. The flex cabling has a very small overall footprint due to the integration of various components in a monolithic product (from top to bottom), with many I/O channels in a row, and combined with a vacuum feed-through and thermalisation blocks. Examples of these components are superconducting sections, IR-filters, low-pass filters, attenuators and couplers, all up to microwave frequencies. By combining various components on flex, Cri/oFlex® is suitable for different quantum computing architectures.

These products gain from standardisation. On one hand, for instance, to harmonise the connectors with other modules, or to set clear and realistic signal, crosstalk, noise, thermal, vacuum, footprint, mechanical and functional requirements. On the other hand, we need to learn requirements from customers so that we can create the products they need and in high volumes as well. Being active in standardisation offers both.

## B.7 INRIM: quantum metrology

INRIM [88] is a public research centre and is Italy's national metrology institute (NMI). INRIM realises, maintains, and develops the national reference standards of the measurement units of the International System (SI), in order to ensure measurements that are reliable and comparable on both a national and international scale.
The INRIM Quantum Metrology and Nanotechnologies (QN) Division handles the mutual application between metrology and topics such as atomic and molecular physics, photonics, quantum electronics, quantum devices and quantum measurements, quantum sensing and enhanced imaging, and actively participates in the Quantum Technologies standardisation initiatives at National, European and International levels, coordinating among the others also EMPIR normative projects focused on QT ([89]). INRIM implements and coordinates the Italian quantum fibre-optic backbone infrastructure ([90]), with the goal to connect to the European Quantum Communication Infrastructure (EuroQCI) which should result from the EU Quantum Flagship.
INRIM is one of the co-founders of the European Metrology Network for Quantum Technologies (EMN-Q) [15], [23], it is member of the CEN CENELEC FGQT since its foundations, and co-founder member of the Italian UNI-FGQT "mirror group" where Italian FGQT contributions are coordinated.

## B.8 CINI: software research for quantum computing

CINI [91] is the main point of reference for the national academic research activities in the fields of Computer Science and Information Technology in Italy. The Consortium involves 1300+ professors of both Computer Science and Computer Engineering, belonging to 39 public universities. In very strict cooperation with the national scientific communities, the Consortium promotes and coordinates scientific activities of research and technological transfer, both basic and applicative, in several fields of Computer Science and Computer Engineering. CINI supports:
- Joint research activities with Universities, Institutes of higher education, research institutions, industries, and Public Administrations.
- The access and the participation in projects and scientific activities of research and technology transfer.
- The creation and development of national research Labs.
- Customised higher education tracks.

Regarding quantum technologies, within CINI's HPC-KTT Laboratory [92], a focus group on quantum computing has been recently established. CINI is also very interested in the standardisation aspects of quantum software. For this reason, a CINI delegate participates in FGQT from its very beginning.

## B.9 Quantumalta: quantum communication

The Quantumalta [93] group within the University of Malta has researchers working on Quantum Communication, mainly QKD as well as Quantum Computing. The QKD team is working on contributing to the standardisation of the software side of QKD components. The Quantumalta group is also active on the physical side of Quantum communication where a quantum link between Malta and Sicily [94] over the telecommunications network has been established which has been subsequently extended to a round-trip between Malta and Sicily [95].

## B.10 Italtel: QKD for secure communications

Italtel [96] is a multinational ICT company that combines the traditional business in networks and communications services with the ability to innovate and develop solutions and applications for digital transformation. Italtel designs and offers end-to-end solutions that address key issues for productivity and business success as well as for the evolution and simplification of network infrastructures. System Design and System Integration are core competencies of Italtel. System Design for us means close interaction with our customer for clearly understanding their needs and goals, tuning to a common language, bringing to the design of the global system and its specific components. Moreover, Italtel has strong System Integration competencies including software development of the components needed for the actual integration of the overall end-to-end system.

Italtel's commitment in quantum technologies is in the design and development of future networks exploiting quantum capabilities for better communication and services. We recognize Quantum Key Distribution (QKD) as the enabling technology for secure communications. Our work is to design geographical networks for the distribution of unconditionally secure encryption keys using QKD as the enabling technology and to integrate it into business applications.

Standardisation plays a strategic role because in a production network we shall integrate and orchestrate devices made by different vendors with different performance levels and set of features. Clear feature, functional and performance description of each node is paramount important for creating an effectively working network. For us having good standards is definitely a must for moving quantum technologies out of the labs.

## B.11 National Physical Laboratory: quantum metrology and standards coordination

The National Physical Laboratory (NPL) [97] is the UK's National Measurement Institute and a world-leading centre of excellence in developing and applying the most accurate measurement standards, science and technology available. For more than a century, NPL has developed and maintained the nation's primary measurement standards. These standards underpin an infrastructure of traceability throughout the UK and the world. NPL employs over 700 scientists, based in south-west London, in a laboratory, which is amongst the world's most extensive and sophisticated measurement science buildings. NPL's work on standards for quantum technologies is done by the Quantum Metrology Institute, which provides measurement standards for the emerging quantum technology industry. The QMI brings together NPL's leading-edge quantum science and metrology research and provides the expertise and facilities needed to test, validate and commercialise new quantum research and technologies.

The QMI is heavily focused on developing the metrology needed for the emerging quantum industry. This includes developing quantitative measurement methods for future measurements services in areas such as computing, communication, sensing and timing as well as establishing capabilities for certification/validation of quantum technology and devices. The work within the QMI is done in collaboration with academia and industry as well as with international partners; and also includes pre-normative research for future international standards which makes the work done within the FGQT highly relevant. The QMI works closely with the British Standards Institute (BSI) and its researches represents the UK's interests in quantum technology and serve as experts within a number of standards developing organisations (SDOs) such as ISO, ETSI an CEN-CENELEC, including the FGQT.

## B.12 Universidad Politécnica de Madrid: quantum networking

Universidad Politécnica de Madrid (UPM, [98]) is the largest Spanish technological university. With two recognitions as Campus of International Excellence, it is outstanding in its research activity together with its training of highly-qualified professionals. It is among the Spanish universities with the greatest research activity and first in the capture of external resources in a competitive regime. UPM signs around 600 contracts with private businesses annually, due to its traditional and close relationship with the industrial and business sectors in all Engineering fields.

Due to its industrial and tech transfer activities, UPM has been involved in QKD standardisation since the very beginning. It was a founding member of the ETSI Industry Specification Group on QKD in 2008, where it chaired (rapporteur) four approved specifications, mostly about quantum network interfaces. It has also participated in the ITU-T activities, both in the Study Group 13 on future networks and in the Focus Group for Quantum Information Technologies for Networks (FGQIT4N). Currently it also participates in the CEN-CENELEC FGQT and is part of the Quantum activities at UNE (the official Spanish Standardisation body). UPM also coordinates the National program on Quantum Communications.

## B.13 QuiX Quantum: photonic quantum computing

QuiX Quantum (Enschede, Netherlands) [99] is a quantum computing company based in Enschede, the Netherlands. The goal of QuiX Quantum is to build a photonic quantum computer. The core product is a quantum photonic processor (QPP), which is currently available with 20 modes, i.e. the world's largest QPP [100]. QuiX Quantum processors are based on Silicon Nitride TriPleX™ waveguide technology, which provides our systems with very low photon loss.

QuiX Quantum is part of the expert group on standardisation of NEN (Dutch standardisation institute), through which we are contributing to the FGQT (focus group on quantum technology) of CEN-CENELEC. The roadmap that is being developed by the FGQT sets out a path to standardisation of quantum technologies, where QuiX Quantum contributed frequently on the topic of photonic quantum computing. The existence of the roadmap is crucial for good standardisation, so that standards are created at the right moment to enable technological progress, but gives no limitations to progress. The importance of standardisation is very high for QuiX Quantum. Standardisation leads to better connectivity and compatibility of the software and hardware interfaces of the products.

## B.14 Orange Quantum Systems: characterising quantum devices

Orange Quantum Systems (Delft, Netherlands) [101] is a systems integrator that is tackling the calibration & characterization bottleneck of (solid-state) quantum processing units. Our products range from customised experiment control software to full-stack quantum characterization systems. Orange Quantum Systems is part of Quantum Delft's ecosystem for innovation of quantum technologies. There are two main areas within quantum technologies where standardisation would greatly benefit our organisation.

As a company focusing on providing the tools to rapidly and reliably characterise quantum devices, we require commonly adopted (standardised) characterization protocols that provide clearly defined, relevant, and explainable metrics. For this purpose, we rely on advanced characterization protocols such as randomised benchmarking and gate set tomography. However, these techniques have been developed for ideal two-level systems (qubits). The assumptions that are required for these protocols are typically violated in physical implementations of quantum systems. Because of this, it requires expert knowledge of both the underlying physics as well as of the protocols to determine what protocol to use, if a protocol needs a custom modification, and to be able to interpret the reported metrics.

As a systems integrator, we need to integrate different components from different manufacturers in our products. Currently the interfaces between these components are not well defined, providing a clear opportunity for standardisation efforts.

## B.15 CNR: quantum technologies

The Italian National Research Council (CNR) [102] is the largest research institution in Italy, funded in 1923. Within CNR, actively involved in the field of Quantum Technologies is the department of Physical Sciences and Technologies of Matter, which started a national program for Quantum Science and Technologies, dedicated in particular to the development of quantum co-processors based on atoms and photons. This will involve integrated packaging systems and microelectronics components as well as the study and characterisation of new nano-structured materials, and infrastructure for the fabrication of semiconductor and superconductor devices, quantum measurements at cryogenic temperatures, devices for the development of quantum computers and sensors, photonics and optoelectronics.

## B.16 NCSR Demokritos: quantum technologies

The National Centre of Scientific Research "Demokritos" (NCSRD, [103]) is a self-governing research organization under the supervision of the Greek Government (the biggest research centre in applied sciences and engineering in Greece). A new Quantum Institute in NCSRD has been established to promote excellence in quantum communications, quantum informatics and relevant technologies at a world-class level. It aspires to become the Greek pole for the advancement of quantum communication and computing and AI-powered quantum research and technology, offering in parallel an effective link to private actors and a wider ecosystem of stakeholders in Greece and the world. NCSRD is participating in the Greek QCI network (HellasQCI) utilizing its mathematical and modelling knowledge and technical expertise regarding QKD systems.

NCSRD' Spin Off company Syndesis Ltd (Syndesis, [104]) is leading the Standardization Working Group and the National Chapter of QuIC for Greece. Syndesis Ltd is utilizing quantum technologies to bring HealthCare industry into the Post-Quantum Era.

## B.17 European Commission: valorising research results and requesting harmonised standards

The European Union has a long tradition of excellence in quantum research. The EU entered in a phase to develop a solid industrial base to valorise the research results. To meet this challenge, the Quantum Technologies Flagship [105] was launched in 2018 and the European High Performance Computing Joint Undertaking (EuroHPC JU) [106] to build state-of-the-art pilot quantum computers by 2023. All 27 MS, with the Commission and with the support of the European Space Agency, towards the development of a quantum communication infrastructure covering the whole EU (EuroQCI) [107].

In 2020, the Commission supported the formation of the CEN-CENELEC Focus Group on Quantum Technologies (FGQT), with the aim of the group being to develop a European roadmap on standardization of Quantum Technology. The Commission has proposed several standardisation activities in quantum research as candidate actions for the 2023 annual EU work programme for European standardisation in support of the EU policy on Quantum Technologies.

The Joint Research Center of the European Commission (JRC), [108], studies developments and trends in quantum technology application areas such as secure communications, computing, simulation, sensing and timing and internal security. While quantum safe communication has a direct contribution to the societal safety and security, using quantum computing resources will also provide new opportunities and capacity to the challenge of encryption within criminal investigations.

# List of abbreviations

| | |
|---|---|
| CEN | European Committee for Standardization |
| CENELEC | European Electrotechnical Committee for Standardization |
| EC | European Commission |
| EMN-Q | European Metrology Network for Quantum Technologies |
| ETSI | European Telecommunications Standards Institute |
| EURAMET | European Association of National Metrology Institutes |
| FG-QIT4N | ITU-T Focus Group on Quantum Information Technology for Networks |
| FGQT | CEN-CENELEC Focus Group on Quantum Technologies |
| GNSS | Global Navigation Satellite System |
| NSB | National Standardisation Body |
| IEC | International Electrotechnical Commission |
| IEEE | Institute of Electrical and Electronics Engineers |
| IETF | Internet Engineering Task Force |
| IRTF | Internet Research Task Force |
| ISG | ETSI Industry Specification Group |
| ISO | International Organization for Standardization |
| ITU-T | International Telecommunication Union - Telecommunications sector |
| JTC | Joint Technical Committee (ISO/IEC, CEN-CENELEC) |
| NMI | European National Metrology Institutes |
| NV | Nitrogen-Vacancy center |
| PAR | Project Authorization Request |
| QED-C | U.S. Quantum Economic Development Consortium |
| QIN | Quantum Information Networks |
| QIRG | Quantum Internet Research Group |
| QIT | Quantum Information Technology |
| QKD | Quantum Key Distribution |
| QKDN | Quantum Key Distribution Networks |
| QMSI | Quantum Metrology & Sensing and quantum enhanced Imaging |
| QT | Quantum Technology |
| QuIC | European Quantum Industry Consortium |
| RFC | Request for Comments |
| SC | ISO/IEC Sub-Committee |
| SDO | Stands Developing Organisation |
| SDQC | Software-Defined Quantum Communication |
| SG | ITU-T Study Group |
| SQL | Standard Quantum Limit |
| TCP/IP | Transmission Control Protocol / Internet Protocol |
| TRL | Technology Readiness Level |

# Declarations

## Ethical Approval and Consent to participate

Not applicable

## Consent for publication

All authors consent for publication.

## Availability of supporting data

Not applicable

## Competing interests

Not applicable

# Funding

| | |
|---|---|
| Horizon 2020 QFlag–CSA, | 820350 |
| Horizon 2020 OpenSuperQ, | 820363 |
| Horizon 2020 Qombs, | 820419 |
| Horizon 2020 CiViQ, | 820466 |
| Horizon 2020 QUANGO, | 101004341 |
| EMPIR MeTISQ, | 19NRM06 |
| EMPIR QADeT, | 20IND05 |
| EMPIR Quantum, | 19NET02 |
| Horizon Europe StandICT.eu 2023, | 951972 |
| UK National Quantum Technologies Programme, | UK department for Business Energy and Industrial Strategy (BEIS) |
| Stichting Quantum Delta Netherlands, | Groeifonds |

# Authors' contributions

All authors contributed to the article

# Acknowledgements

Parts of the work were funded by European Union's Horizon 2020 and Horizon Europe research and innovation program under the projects QFlag-CSA (Grant No. 820350), QUCATS-CSA (Grant No. 101070193), OpenSuperQ project (Grant Nr. 820363), Qombs (Grant No. 820419), CiViQ (Grant No. 820466) and QUANGO (Grant No. 101004341).

Part of the work was funded by the projects EMPIR 19NRM06 MeTISQ, 20IND05 QADeT and 19NET02 Quantum - these projects received funding from the EMPIR program co-financed by the Participating States and from the European Union Horizon 2020 research and innovation program.

Part of the work was funded by four individual subgrants from Horizon Europe StandICT.eu 2023 (Grant No. 951972).

Part of the work was funded by the UK department for Business Energy and Industrial Strategy (BEIS) through the UK National Quantum Technologies Programme.

Part of the work was funded by a grant from the Quantum Delta Netherlands Foundation.

# References

[1]     Blind, Knut, "Standardisation as a Catalyst for Innovation. Erasmus Research Institute of Management (ERIM)", Erasmus University and the Erasmus School of Economics (ESE), https://ideas.repec.org/p/ems/euriar/17558.html, 2009.

[2]     Standards developing organisations, https://www.hse.gov.uk/comah/sragtech/docspubstand.htm

[3]     Jenet, A., Trefzger, C., Lewis, A.M., Taucer, F., Van Den Berghe, L., Tüchler, A., Loeffler, M. and Nik, S. (2020). Standards4Quantum: Making Quantum Technology Ready for Industry - Putting Science into Standards, EU Publications Office, Luxembourg, doi:10.2760/882029, https://publications.jrc.ec.europa.eu/repository/handle/JRC118197.

[4]     Technology Readiness Level, https://en.wikipedia.org/wiki/Technology_readiness_level

[5]     Website of the QFlag – Quantum Flagship Coordination and Support Action, funded by the European Commission. online (22.02.2022), https://qt.eu/about-quantum-flagship/introduction-to-the-quantum-flagship.

[6]     Quantum Flagship, Strategic Research Agenda, https://qt.eu/app/uploads/2020/04/Strategic_Research-_Agenda_d_FINAL.pdf, April 2020

[7]     CEN-CENELEC Focus Group on Quantum Technologies, FGQT, https://www.cencenelec.eu/areas-of-work/cen-cenelec-topics/quantum-technologies/

[8]     IEEE/Open Group 1003.1-2017, "Standard for Information Technology—Portable Operating System Interface (POSIX(TM)) Base Specifications", Issue 7, https://ieeexplore.ieee.org/document/8277153, 2017

[9]     IBM, "OpenQASM 3.x Live Specification", https://qiskit.github.io/openqasm/index.html, 2022

[10]    A. Dahlberg et al., "NetQASM -- A low-level instruction set architecture for hybrid quantum-classical programs in a quantum internet", https://arxiv.org/abs/2111.09823, 2021

[11]    Dowling, J.P.; Milburn, G.J. Quantum technology: The second quantum revolution. Philos. Trans. R. Soc. A 2003, 361, 1655–1674. https://royalsocietypublishing.org/doi/pdf/10.1098/rsta.2003.1227

[12]    Giovannetti, V.; Maccone, L.; Lloyd, S. Quantum-Enhanced Measurements: Beating the Standard Quantum Limit. Science 306, Issue 5700, pp. 1330-1336 (2004). https://www.science.org/doi/10.1126/science.1104149

[13]    I. Ruo Berchera, I. P Degiovanni. Quantum imaging with sub-Poissonian light: challenges and perspectives in optical metrology. Metrologia 56 024001 (2019). https://iopscience.iop.org/article/10.1088/1681-7575/aaf7b2/pdf

[14]    Giovannetti, V., Lloyd, S. & Maccone, L. Advances in quantum metrology. Nature Photon 5, 222–229 (2011). https://doi.org/10.1038/nphoton.2011.35

[15]    European Metrology Network for Quantum Technologies, www.euramet.org/quantum-technologies

[16]    White Paper On Quantum Metrology & Sensing And Quantum Enhanced Imaging, ZEISS SYMPOSIUM 2018.

https://www.zeiss.com/content/dam/Corporate/innovation_and_technology/downloads/zeiss-symposium_whitepaper_qms_qei.pdf

[17] Quantum Flagship, "New Strategic Research Agenda on Quantum technologies", 2022 https://digital-strategy.ec.europa.eu/en/news/new-strategic-research-agenda-quantum-technologies

[18] Antonio Acín et al 2018 New J. Phys. 20 080201, https://iopscience.iop.org/article/10.1088/1367-2630/aad1ea

[19] Bureau International des Poids et Measures, "The International System of Units (SI)", https://www.bipm.org/en/measurement-units

[20] Bureau International des Poids et Measures, "BIPM Workshop: The Quantum Revolution in Metrology", https://www.bipm.org/en/bipm-workshops/quantum-metrology

[21] QuantERA, "Transnational co-funded Calls for Proposals", https://quantera.eu/quantera-funded-projects/

[22] European Quantum Flagship, "Learn more about Quantum projects", https://qt.eu/about-quantum-flagship/projects/

[23] I.P. Degiovanni, M. Gramegna, S. Bize, H. Sherer, C.J. Chunilall. EURAMET EMN-Q: The European metrology network for quantum technologies". Measurement: Sensors, Volume 18, 100348 (2021). https://www.sciencedirect.com/science/article/pii/S2665917421003111

[24] C. H. Bennett, G. Brassard, "Quantum cryptography: public key distribution and coin tossing",Proceedings of IEEE international conference on computers, systems and signal processing. vol. 175. New York. 1984. p. 8, https://arxiv.org/abs/2003.06557, DOI:10.1016/j.tcs.2011.08.039

[25] N. Gisin, G. Ribordy, W. Tittel, H. Zbinden. Quantum cryptography. Rev Mod Phys. 2002;74:145–95. DOI: https://doi.org/10.1103/RevModPhys.74.145

[26] V. Scarani, H. Bechmann-Pasquinucci, N. J. Cerf, M. Dušek, Norbert Lütkenhaus, M. Peev. The security of practical quantum key distribution. Rev Mod Phys. 2009;81:1301-1350, DOI: https://doi.org/10.1103/RevModPhys.

[27] M. Peev et al., The SECOQC quantum key distribution network in Vienna. NJP 2009: 11, 075001, DOI: https://doi.org/10.1088/1367-2630/11/7/075001

[28] V. Martin et al., Quantum technologies in the telecommunications industry. EPJ Quantum Technology 2021: 8, 19, DOI: https://doi.org/10.1140/epjqt/s40507-021-00108-9

[29] S. Wehner, D. Elkouss, R. Hanson. Quantum internet: A vision for the road ahead, Science 2018: 362(6412) DOI: https://www.science.org/doi/10.1126/science.aam9288

[30] OIDA, "OIDA Quantum Photonics Roadmap: Every Photon Counts," Optica Industry Report, 3 (2020). https://opg.optica.org/abstract.cfm?URI=OIDA-2020-3

[31] B. Weder, J. Barzen, F. Leymann, and M. Zimmermann, "Hybrid quantum applications need two orchestrations in superposition: A software architecture perspective", in 2021 IEEE International Conference on Web Services (ICWS). IEEE, 2021, pp. 1–13, https://www.iaas.uni-stuttgart.de/publications/Weder2021_OrchestrationsInSuperposition.pdf

[32] CEN-CENELEC FGQT, "Standardisation Roadmap for Quantum Technologies", document N020, latest version via https://www.cencenelec.eu/areas-of-work/cen-cenelec-topics/quantum-technologies/

[33] ETSI, "Industry specification group (ISG) on quantum key distribution for users (qkd)", https://www.etsi.org/committee/1430-qkd

[34] ETSI GS QKD 002, "Quantum Key Distribution; Use Cases", https://www.etsi.org/deliver/etsi_gs/qkd/001_099/002/01.01.01_60/gs_qkd002v010101p.pdf, June 2010.

[35] ETSI GR QKD 007, "Quantum Key Distribution (QKD); Vocabulary", https://www.etsi.org/deliver/etsi_gr/QKD/001_099/007/01.01.01_60/gr_qkd007v010101p.pdf, December 2018.

[36] ETSI GS QKD 004, "Quantum Key Distribution (QKD); Application Interface", https://www.etsi.org/deliver/etsi_gs/QKD/001_099/004/02.01.01_60/gs_qkd004v020101p.pdf, August 2020.

[37] ETSI GS QKD 014, "Quantum Key Distribution (QKD); Protocol and data format of REST-based key delivery API", https://www.etsi.org/deliver/etsi_gs/QKD/001_099/014/01.01.01_60/gs_qkd014v010101p.pdf, February 2019.

[38] ETSI GS QKD 015, "Quantum Key Distribution (QKD); Control Interface for Software Defined Networks", https://www.etsi.org/deliver/etsi_gs/QKD/001_099/015/01.01.01_60/gs_QKD015v010101p.pdf, March 2021.

[39] Marco Lucamarini, Andrew Shields, Romain Alléaume, Christopher Chunnilall, Ivo Pietro Degiovanni, Marco Gramegna, Atilla Hasekioglu, Bruno Huttner, Rupesh Kumar, Andrew Lord, Norbert Lütkenhaus, Vadim Makarov, Vicente Martin, Alan Mink, Momtchil Peev, Masahide Sasaki, Alastair Sinclair, Tim Spiller, Martin Ward, Catherine White, Zhiliang Yuan, "Implementation Security of Quantum Cryptography Introduction, challenges, solutions", ETSI White Paper No. 27, ISBN No. 979-10-92620-21-4, https://www.etsi.org/images/files/ETSIWhitePapers/etsi_wp27_qkd_imp_sec_FINAL.pdf, July 2018.

[40] ETSI GS QKD 005, "Quantum Key Distribution (QKD); Security Proofs", https://www.etsi.org/deliver/etsi_gs/qkd/001_099/005/01.01.01_60/gs_qkd005v010101p.pdf, December 2010.

[41] ETSI GS QKD 008, "Quantum Key Distribution (QKD); QKD Module Security Specification", https://www.etsi.org/deliver/etsi_gs/qkd/001_099/008/01.01.01_60/gs_qkd008v010101p.pdf, December 2010

[42] ETSI GR QKD 003, "Quantum Key Distribution (QKD); Components and Internal Interfaces", https://www.etsi.org/deliver/etsi_gr/QKD/001_099/003/02.01.01_60/gr_qkd003v020101p.pdf, March 2018.

[43] ETSI GS QKD 011, "Quantum Key Distribution (QKD); Component characterization: characterizing optical components for QKD systems", https://www.etsi.org/deliver/etsi_gs/qkd/001_099/011/01.01.01_60/gs_qkd011v010101p.pdf, May 2016.

[44] ISO/IEC 15408, "Evaluation criteria for IT security", https://www.noraonline.nl/wiki/ISO/IEC_15408, 2010-2011

| | |
|---|---|
| [45] | ITU-T Study Group 13, "Future networks, with focus on IMT-2020, cloud computing and trusted network infrastructure", https://www.itu.int/en/ITU-T/about/groups/Pages/sg13.aspx |
| [46] | ITU-T Study Group 17, "Security",https://www.itu.int/en/ITU-T/about/groups/Pages/sg17.aspx |
| [47] | ITU-T Y.3800, "Overview on networks supporting quantum key distribution", https://www.itu.int/itu-t/recommendations/rec.aspx?rec=13990, October 2019. |
| [48] | ITU-T Y.3801, "Functional requirements for quantum key distribution networks", https://www.itu.int/itu-t/recommendations/rec.aspx?rec=14258, April 2020. |
| [49] | ITU-T Y.3802,, "Quantum key distribution networks – Functional architecture", https://www.itu.int/itu-t/recommendations/rec.aspx?rec=14407, December 2020. |
| [50] | ITU-T Y.3803 (12/2020), "Quantum key distribution networks – Key management", https://www.itu.int/itu-t/recommendations/rec.aspx?rec=14408, December 2020. |
| [51] | ITU-T Y.3804, "Quantum key distribution networks – Control and management", https://www.itu.int/itu-t/recommendations/rec.aspx?rec=14409, September 2020. |
| [52] | ITU-T X.1710 (10/2020), "Security framework for quantum key distribution networks", https://www.itu.int/itu-t/recommendations/rec.aspx?rec=14452, October 2020. |
| [53] | ITU-T X.1714 (10/2020), "Key combination and confidential key supply for quantum key distribution networks", https://www.itu.int/itu-t/recommendations/rec.aspx?rec=14453, October 2020. |
| [54] | ITU-T X.1702 (11/2019), "Quantum noise random number generator architecture", https://www.itu.int/itu-t/recommendations/rec.aspx?rec=14095, November 2019. |
| [55] | ITU-T Focus Group on Quantum Information Technology for Networks (FG-QIT4N), https://www.itu.int/en/ITU-T/focusgroups/qit4n/Pages/default.aspx |
| [56] | ITU-T Focus Group Publications, https://www.itu.int/pub/T-FG |
| [57] | ISO/IEC JTC1 WG14 on Quantum Computing, https://jtc1info.org/technology/working-groups/quantum-computing/ |
| [58] | ISO/IEC AWI 4879, "Information technology — Quantum computing — Terminology and vocabulary", https://www.iso.org/standard/80432.html, https://bds-bg.org/en/project/show/iso:proj:80432, under development, February 2022. |
| [59] | ISO/IEC JTC 1/SC 27/WG 3, "Security evaluation, testing and specification, https://standards.iteh.ai/catalog/tc/iso/56ffc1fc-b504-40a6-b4ab-3cacf8ff9f7d/iso-iec-jtc-1-sc-27-wg-3 |
| [60] | NEN-EN-ISO/IEC 15408-1, "Evaluation criteria for IT security - Part 1: Introduction and general model", https://www.nen.nl/nen-en-iso-iec-15408-1-2020-en-269562, 2020 |
| [61] | ISO/IEC CD 23837-1.2, "Security requirements, test and evaluation methods for quantum key distribution — Part 1: Requirements", https://www.iso.org/standard/77097.html, under development, February 2020. |
| [62] | IEEE, "Quantum Initiative Support for Standards", https://quantum.ieee.org/standards |
| [63] | IEEE, "Developing standards", https://standards.ieee.org/develop/index.html |
| [64] | IEEE P1913, "Software-Defined Quantum Communication", https://standards.ieee.org/project/1913.html |
| [65] | IEEE P7130, "Standard for Quantum Technologies Definitions", https://standards.ieee.org/project/7130.html |
| [66] | IEEE P7131, "Standard for Quantum Computing Performance Metrics & Performance Benchmarking", https://standards.ieee.org/project/7131.html |
| [67] | IEEE P3120, "Standard for Quantum Computing Architecture", https://standards-dev21.ieee.org/ieee/3120/10751/ |
| [68] | Quantum Internet Research Group (QIRG), https://irtf.org/qirg |
| [69] | Quantum Internet Research Group (QIRG) Mailing List https://www.irtf.org/mailman/listinfo/qirg |
| [69] | Quantum Internet Research Group (QIRG) Mail Archive, https://mailarchive.ietf.org/arch/browse/qirg/ |
| [70] | Quantum Internet Research Group (QIRG) Documents, https://datatracker.ietf.org/rg/qirg/documents/ |
| [71] | Kozlowski, W., Wehner, S., Van Meter, R., Rijsman, B., Cacciapuoti, A., Caleffi, M., Nagayama, S., "Architectural Principles for a Quantum Internet", Work in Progress, Internet-Draft https://datatracker.ietf.org/doc/draft-irtf-qirg-principles/, 14 February 2022. |
| [72] | Wang, C., Rahman, A., Li, R., Aelmans, M., Chakraborty, K., "Application Scenarios for the Quantum Internet", Work in Progress, Internet-Draft https://datatracker.ietf.org/doc/draft-irtf-qirg-quantum-internet-use-cases/, 20 August 2021. |
| [73] | Namazi, M., "Practical Quantum Networking at Room Temperature", https://www.youtube.com/watch?v=2ELYL71tlD8, 26 March 2021. |
| [74] | Dür, W., "Genuine and Optimized Entanglement-Based Quantum Networks" https://www.youtube.com/watch?v=j-Ri-RRfUXY, 23 September 2021. |
| [75] | Kaplan, M., "Building Quantum Networks at the Local-Area Scale", https://www.youtube.com/watch?v=D_Nb43-uicY, 3 February 2022. |
| [76] | Euramet, The gateway to Europe's integrated metrology community, https://www.euramet.org/ |
| [77] | QuIC, "European Quantum Industry Consortium", https://www.euroquic.org/ |
| [78] | StandICT, "Supporting European Experts Presence in International Standardisation Activities in ICT", https://standict.eu/about, online 1.2.2022 |
| [79] | TNO, The Netherlands Organisation for Applied Scientific Research, https://tno.nl |
| [80] | Quantum technologies at TNO, https://www.tno.nl/en/focus-areas/industry/roadmaps/semiconductor-equipment/quantum-technology/ |
| [81] | Horizon 2020 QFlag Coordination and Support Action, https://qt.eu/about-quantum-flagship/introduction-to-the-quantum-flagship/qflag-quantum-flagship-coordination-and-support-action/, https://cordis.europa.eu/project/id/820350, 2019-2022. |
| [82] | Horizon 2020 QUCATS Coordination and Support Action, https://qt.eu/about-quantum-flagship/projects/qucats/, 2022-2025. |
| [83] | Quantum Delta Netherlands, https://quantumdelta.nl/ |
| [84] | PTB, "German National Metrology Institute", https://www.ptb.de/cms/ |
| [85] | DIN, "German institute for Standardisation", https://www.din.de/en |
| [86] | Huawei Dusseldorf, https://www.huawei.com/de/impressum |
| [87] | Delft Circuits, "Hardware for quantum engineers", https://delft-circuits.com/, |
| [88] | INRIM, the Italian National Metrology Institute, https://www.inrim.it |

[89] EMPIR Normative project Use Case: the 19NRM06 MeTISQ project, https://www.euramet.org/project-19nrm06
[90] In-field tests and calibrations over the I-QB Italian Quantum Backbone, https://phys.org/news/2022-01-technique-long-distance-quantum-key-real-world.html
[91] CINI, the National Interuniversity Consortium for Informatics https://www.consorzio-cini.it/index.php/en/
[92] HPC: Key Technologies and Tools https://www.consorzio-cini.it/index.php/en/national-laboratories/hpc-key-technologies-and-tools
[93] Quantumalta, University of Malta, https://quantum.edu.mt/
[94] Sören Wengerowsky et al, "Entanglement distribution over a 96-km-long submarine optical fiber", PINAS, 2 Apr 2019, https://www.pnas.org/content/116/14/6684
[95] Sören Wengerowsky et al, "Passively stable distribution of polarisation entanglement over 192 km of deployed optical fibre", npj Quantum Information, 10 Jan 2020, https://www.nature.com/articles/s41534-019-0238-8
[96] Italtel, "Italian telecommunications equipment and ICT", https://www.italtel.com/
[97] NPL, "Creating impact from science and engineering", https://www.npl.co.uk/
[98] Universidad Politécnica de Madrid, https://www.upm.es/
[99] QuiX Quantum, "The fastest way to a quantum future", https://www.quixquantum.com/
[100] Caterina Taballione et al., "20-mode Universal Quantum Photonic Processor", to be submitted, https://quantumbusinessnetwork.de/en/events/quix-quantum-webinar-20-mode-quantum-photonic-processor/, 3 March 2022.
[101] Orange Quantum Systems, "Solutions to tackle the quantum device characterization bottleneck", https://orangeqs.com/
[102] CNR, the Italian National Research Council, https://www.cnr.it/en
[103] NCSRD, the National Centre of Scientific Research "Demokritos", https://www.demokritos.gr/
[104] Syndesis Ltd, "HealthCare in the Post-Quantum Era", https://www.syndesis.eu/
[105] European Quantum Technology Flagship, https://qt.eu/
[106] The European High Performance Computing Joint Undertaking (EuroHPC JU), https://eurohpc-ju.europa.eu/index_en
[107] The European Quantum Communication Infrastructure (EuroQCI) Initiative, https://digital-strategy.ec.europa.eu/en/policies/european-quantum-communication-infrastructure-euroqci
[108] European Commission – Joint Research Center, https://ec.europa.eu/info/departments/joint-research-centre_en